\documentclass[fleqn,usenatbib]{mnras}

\usepackage[T1]{fontenc}

\DeclareRobustCommand{\VAN}[3]{#2}
\let\VANthebibliography\thebibliography
\def\thebibliography{\DeclareRobustCommand{\VAN}[3]{##3}\VANthebibliography}

\usepackage{graphicx}	% Including figure files
\usepackage{amsmath}	% Advanced maths commands
\usepackage{amssymb}	% Extra maths symbols
\usepackage{verbatim}
\usepackage{physics}
\usepackage{xspace}
\usepackage{natbib}
 
\usepackage{newtxtext,newtxmath}
\newcommand\ABACUS{\textsc{abacus}\xspace}
\newcommand\GADGET{\textsc{gadget-2}\xspace}

\newcommand\SWIFT{\textsc{swift}\xspace}
\newcommand\ROCKSTAR{\textsc{rockstar}\xspace}
\newcommand\FASTPM{\textsc{fastpm}\xspace}
\title[The DESI $N$-body Simulation Code Comparison Project]{The DESI $N$-body Simulation Project I: Testing the Robustness of Simulations for the DESI Dark Time Survey}%\\
\author[C. Grove et al.]{Cameron Grove$^{1}$\thanks{E-mail: cameron.grove@durham.ac.uk},
Chia-Hsun Chuang$^{2,3}$\thanks{E-mail: albert.chuang@utah.edu},
Ningombam Chandrachani Devi$^{4,5}$,
\newauthor
Lehman Garrison$^{6}$,
Benjamin L'Huillier$^{7}$,
Yu Feng$^{8}$,
John Helly$^{1}$,
\newauthor
C\'{e}sar Hern\'{a}ndez-Aguayo$^{9,10}$,
Shadab Alam$^{11}$,
Hanyu Zhang$^{12}$,
Yu Yu$^{13}$,
Shaun Cole$^{1}$,
\newauthor
Daniel Eisenstein$^{14}$,
Peder Norberg$^{1,15}$,
Risa Wechsler$^{2}$,
David Brooks$^{16}$,
\newauthor
Kyle Dawson$^{3}$,
Martin Landriau$^{17}$,
Aaron Meisner$^{18}$,
Claire Poppett$^{19}$,
Gregory Tarl\'e$^{20}$,
\newauthor
Octavio Valenzuela$^{4}$  
\\
\newline
\newline
\emph{\normalsize Affiliations are listed at the end of the paper}
\\
}

\date{Accepted XXX. Received YYY; in original form ZZZ}

\pubyear{2021}

\begin{document}
\label{firstpage}
\pagerange{\pageref{firstpage}--\pageref{lastpage}}
\maketitle

\begin{abstract}
Analysis of large galaxy surveys requires confidence in the robustness of numerical simulation methods.
The simulations are used to construct mock galaxy catalogs to validate data analysis pipelines and identify potential systematics.
We compare three $N$-body simulation codes, \ABACUS, \GADGET, and \SWIFT, to investigate the regimes in which their results agree. We run $N$-body simulations at three different mass resolutions, $6.25\times10^{8}$, $2.11\times10^{9}$, and $5.00\times10^{9}~h^{-1}$M$_{\sun}$, matching phases to reduce the noise within the comparisons.
We find systematic errors in the halo clustering between different codes are smaller than the DESI statistical error for $s > 20\, h^{-1}$Mpc in the correlation function in redshift space. 
Through the resolution comparison we find that simulations run with a mass resolution of $2.1\times10^{9}~h^{-1}$M$_{\sun}$ are sufficiently converged for systematic effects in the halo clustering to be smaller than the DESI statistical error at scales larger than $20 \, h^{-1}$Mpc.
These findings show that the simulations are robust for extracting cosmological information from large scales which is the key goal of the DESI survey.
Comparing matter power spectra, we find
the codes agree to within 1\% for $k \leq 10~h$Mpc$^{-1}$. 
We also run a comparison of three initial condition generation codes
and find good agreement.
In addition, we include a quasi-$N$-body code, FastPM, since we plan use it for certain DESI analyses.
The impact of the halo definition and galaxy-halo relation will be presented in a follow up study.
\end{abstract}


%%%%%%%%%%%%%%%%%%%%%%%%%%%%%%%%%%%%%%%%%%%%%%%%%%

%%%%%%%%%%%%%%%%% BODY OF PAPER %%%%%%%%%%%%%%%%%%

% None of the figures are currently in their final form
% To do list for improving figures:
%
% Update figures taken from cosmosim slides so that they are less confusing
% and more specific: clearly showing exactly what I want them to
%
% Consistent colours for codes between figures
%
% Consistent fonts/labels/styles
%
%
%

\section{Introduction}

For many years $N$-body simulations have been used as a tool to explore the nonlinear evolution of the distribution of matter in the Universe \citep{Davis1985}. Their use has been invaluable in creating mock galaxy catalogs to validate the results of surveys such as 2dFGRS, SDSS, DES, KiDS, and more %\citep{Cole1998,Howlett_2015,Avila2018}. 
\citep{cole1998,Rodriguez-Torres:2015vqa,derose2021dark,deJong2012}.
The Dark Energy Spectroscopic Instrument (DESI) will be performing surveys of unprecedented size over the next five years, measuring tens of millions of galaxy spectra \citep{collaboration2016desi}. The large size of the DESI survey means that the statistical errors on key measured quantities will be small and therefore keen attention must be paid to systematic errors in all stages of the data collection and analysis. Measuring the size of systematic errors introduced by $N$-body simulations is imperative to understand their impact on mock galaxy catalogs and analysis pipelines.

One way to measure systematic errors for an individual $N$-body code is using convergence testing. The code can be run with progressively more accurate parameter choices until measured statistics no longer change \citep{Power_2003,DeRose_2019}. However, convergence testing is limited because it only provides information on the systematic errors which can be reduced by running a code with greater numerical accuracy. 
%There can exist other systematic errors which are present in the converged simulations. 
Different implementations or assumptions adopted by different codes could introduce systematic errors as well.
Comparing the results from several different codes can indicate the level of control over systematic errors in converged runs of $N$-body simulation codes.

Code comparison projects measure the precision of $N$-body simulation codes when run from identical initial conditions. Measurements of the power spectrum, clustering and halo properties can be compared to indicate the precision to which $N$-body simulations can estimate these quantities \citep{Winther_2015,Schneider2016,Garrison2019}.

This paper contains the results from multiple code comparison projects using several modern $N$-body simulation methods. Firstly we compare the initial conditions created using several different codes, measuring the power spectrum along with particle velocity statistics. The main $N$-body code comparison project compares simulations run with \ABACUS, \FASTPM, \GADGET, and \SWIFT from identical initial conditions. The matter power spectrum is measured, along with dark matter halo properties, abundances, clustering and power spectra. The precision of these results between different $N$-body codes and simulation resolutions is measured.

In Section~\ref{sec:ICC}, the initial conditions comparison is performed between three  different codes.  In Section~\ref{sec:Nbody}, we discuss the different codes used to run the $N$-body simulation and we compare the results of the $N$-body simulations in terms of the matter power spectrum. In Section~\ref{sec:Halo} we compare the halo catalogs produced from the different simulation codes, including measurements of the clustering, halo mass functions, and matched halo properties. Finally we summarize and conclude in Section~\ref{sec:Con}.

%\section{Initial Conditions Comparison}
\section{Initial Conditions Codes}
\label{sec:ICC}

%{\color{red} Brief overview of initial conditions in $N$-body simulations and why they're important.

%Here may be a good place to discuss the simulation requirements of DESI which informed the choice of simulations to compare, ie mass resolution \& boxsize necessary for effective use in BAO \& RSD measurements. Table 1 will be in this section, or wherever the reasoning for simulation requirements is discussed.}

The starting points for $N$-body simulations are the initial conditions (ICs). These define the positions, velocities and masses of particles which are to be fed into the $N$-body simulation code. Analysis of the cosmic microwave background radiation suggests that the initial matter density field produced during inflation is a Gaussian random field \citep{Aghanim:2018eyx}. The properties of this field are determined by its power spectrum, which in turn is determined by the cosmology of the universe. Different realisations of the same power spectrum are possible as the phases of the field are randomly generated \citep{Abrahamsen1997}.

%Chani is working on this section.
In this section, we discuss three different IC generators. The considered IC codes include those used in conjunction with the \ABACUS \citep{Abacus_code_paper} and \FASTPM \citep{FASTPM} $N$-body codes, in addition to a combination of two codes, \textsc{panphasia}\footnote{\url{http://icc.dur.ac.uk/Panphasia.php}}\citep{Jenkins2013} and \textsc{ic\_gen} which we refer to as \textsc{panphasia} in this text, see Section~\ref{sec:panphasia} for details. 

The cosmology is set according to the flat $\Lambda$CDM Planck 2018 results \citep{Aghanim:2018eyx} with the following parameters: the Hubble expansion rate at present time, $H_{0} = 67.36$~km~s$^{-1}$, $\Omega_c h^2 = 0.1200$,  $\Omega_b h^2 = 0.0223$,  $\Omega_{\Lambda} = 0.6834$,  r.m.s linear density fluctuation, $\sigma_{8} = 0.8$, scalar spectral index  $n_{s} = 0.9649$.

% don't include A_s for brevity. It is determined by sigma_8
%and the scalar power amplitude $A_{s} = 2.1 \times 10^{-9}$. 

Considering the requirements of the DESI project, we run the simulations in boxes of side length, $L_{\textrm{box}} = 500 h^{-1} {\rm Mpc}$ with three different dark matter particle numbers: $N_{\textrm{p}} = 1296^3$, $ 1728^3$ and $2592^3$. The initial conditions are generated at redshift, $z_{\textrm{ini}}=199$ with the input theoretical matter power spectrum generated using \textsc{camb}\footnote{\url{https://camb.info/}}. The details of the methods employed by each code are described below. We study the consistency between the codes in both  the dark matter position and velocity distributions. The convergence test in each individual code is also performed using the three sets of resolutions.   
In Table~\ref{tab:1}, we provide the specifications of the simulations.

\begin{table}
	\centering
	\caption{The number of particles and particle masses of the simulation boxes. All are cubic boxes with a side length of 500 Mpc/h.}
	\label{tab:1}
	\begin{tabular}{lcr} 
		\hline
		Simulation & Particle Number & Particle Mass /$h^{-1}$M$_{\sun}$ \\
		\hline
		Low Resolution & $1296^3$ & $5.00\times10^{9}$ \\
		Medium Resolution & $1728^3$ & $2.11\times10^{9}$ \\
		High Resolution & $2592^3$ & $6.25\times10^{8}$ \\
		\hline
	\end{tabular}
\end{table}

\subsection{Code details}

This section contains descriptions of the codes used to create initial conditions used in the comparisons in sections \ref{sec:ICPS} and \ref{sec:ICstats}.

\subsubsection{ABACUS IC}

The \ABACUS IC code, called \textsc{zeldovich-plt}\footnote{\url{https://github.com/abacusorg/zeldovich-PLT}}, generates Zel'dovich Approximation initial conditions, optionally applying Particle Linear Theory (PLT) corrections \citep{Marcos2006,Garrison2016} for the breaking of linear theory that occurs on small scales due to the discrete representation of the density field by particles.  The PLT corrections are not enabled for purposes of this code comparison.  The code is written in C++ and is designed to produce initial conditions in memory-limited environments by buffering the state to disk.  The code uses double precision internally, and the positions are output as single-precision displacements on a lattice.  The random number generator for the initial Gaussian density field outputs 64 bits and allows synchronizing the white noise at different resolutions; this capability is employed for the three resolutions of this IC code comparison.  The input power spectrum normalization is recomputed by \textsc{zeldovich-plt} then scaled back to the requested redshift using the growth factor ratio from \ABACUS's cosmology module.  Modes are filled out to the Nyquist sphere.  This is the initial conditions generator used by the \ABACUS code, including the AbacusSummit project\footnote{\url{https://abacussummit.readthedocs.io}} \citep{AbacusSummit}.

\subsubsection{\textsc{ic\_gen} / \textsc{panphasia}}
\label{sec:panphasia}
There are two components to this method of creating the initial conditions which has been labelled \textsc{panphasia}. Firstly \textsc{panphasia} describes a method of setting the phases of cosmological initial conditions. \textsc{panphasia} describes a pseudo random number sequence mapped to a cosmological volume from which subsections can be taken to provide the phases for initial conditions \citep{Jenkins2013}. This allows simple creation of resimulation initial conditions and also for the phases used in an $N$-body simulation to be shared so that others can recreate the same cosmological volume at any resolution without having to share the full phase information.

Once the phases have been generated using \textsc{panphasia} the initial conditions are generated using a proprietary code called \textsc{ic\_gen}. This uses second order Lagrangian perturbation theory (2LPT) \citep{Jenkins2010} to create the initial conditions particle distribution according to the phases specified by the location within \textsc{panphasia}.

\subsubsection{FastPM IC}
While not originally designed as an initial condition code,
FastPM does compute an initial particle state with the traditional 2LPT-on-a-mesh approach.
In addition, the initial condition code in FastPM supports producing constrained Gaussian realizations. 

Particles are first placed on a uniform grid with $N_c$ particles per side,
then the first and second order LPT displacement ($\mathbf{s}_{1,2}$) and velocity terms
($\mathbf{v}{1,2}$) are calculated from a potential 
induced from a Gaussian white noise field $g(\mathbf{x})$ according to an isotropic linear power spectrum $P_\mathrm{lin}(k)$.
The white noise field is sampled on a finite mesh of $B_\mathrm{LPT} N_c$ samples per side
from a pseudo-random generator compatible to the one used by Gadget's N-GenIC\citep{Angulo2012,Springel2005}.

The choice of $B_\mathrm{LPT}$, the finite differentiation operators, and the finite Laplace operator all affect
the initial particle state, even though the effect on the late time non-linear field can be quite minimal.

%\subsection{Code Comparison}
\subsection{Comparison of Initial Conditions}
In this section, we compare the initial conditions in terms of particle and velocity distributions.
%\subsection{Power spectrum Comparison}
\subsubsection{Comparison of Particle Distribution}
\label{sec:ICPS}
The power spectrum is a common measure for analysing both initial conditions and evolved $N$-body simulations. It measures the strength of the matter density contrast at different scales.

The power spectrum, $P(k)$ is defined as the Fourier transform of the autocorrelation function, $\xi(r)$ which in turn is derived from the matter overdensity, $\delta(\mathbf{x}) = \frac{\rho(\mathbf{x)}}{\bar{\rho}} - 1 $,

\begin{equation}
    P(k) = \int{}d^3\mathbf{r}\ e^{-i\mathbf{k}\cdot\mathbf{r}}\xi(r),
\label{eq:pk_for}    
\end{equation}

\begin{equation}
    \xi(r) = \langle\delta(\mathbf{x})\delta(\mathbf{x} - \mathbf{r})\rangle = \frac{1}{V} \int{}d^{3}\mathbf{x}\ \delta(\mathbf{x})\delta(\mathbf{x} - \mathbf{r}).
\label{eq:corr_fun}    
\end{equation}

Initial conditions sample the matter density field producing a set of particles. This limited number of particles means that there is a maximum $k$ above which the power spectrum cannot be represented in the initial conditions. This is the Nyquist frequency. For a box with sidelength $L$ and $N^{3}$ particles the wavenumber associated with the Nyquist frequency is

\begin{equation}
    k_{\rm Nyquist} = \frac{\pi N}{L}.
\end{equation}
The Nyquist wavenumber, and therefore the Nyquist frequency, increases with number of particles allowing us to compare to the reference power spectrum at higher $k$.

The power spectra are measured using \textsc{nbodykit} \footnote{https://nbodykit.readthedocs.io/en/latest/} \citep{Hand2018} using a triangular shaped cloud mass assignment scheme and grid sizes of $2592^{3}$, $3456^{3}$ $5184^{3}$ were used for the low, medium and high resolution ICs respectively. Differences near the Nyquist frequency are observed when measuring the power spectra with different grid sizes and  aliasing is observed when using a grid size which does not evenly divide the particle distribution.  Using large grid sizes which are multiples of the particle numbers mitigates these effects. The power spectra are truncated at the Nyquist frequency for each resolution.

\begin{figure*}
	\includegraphics[width=\textwidth]{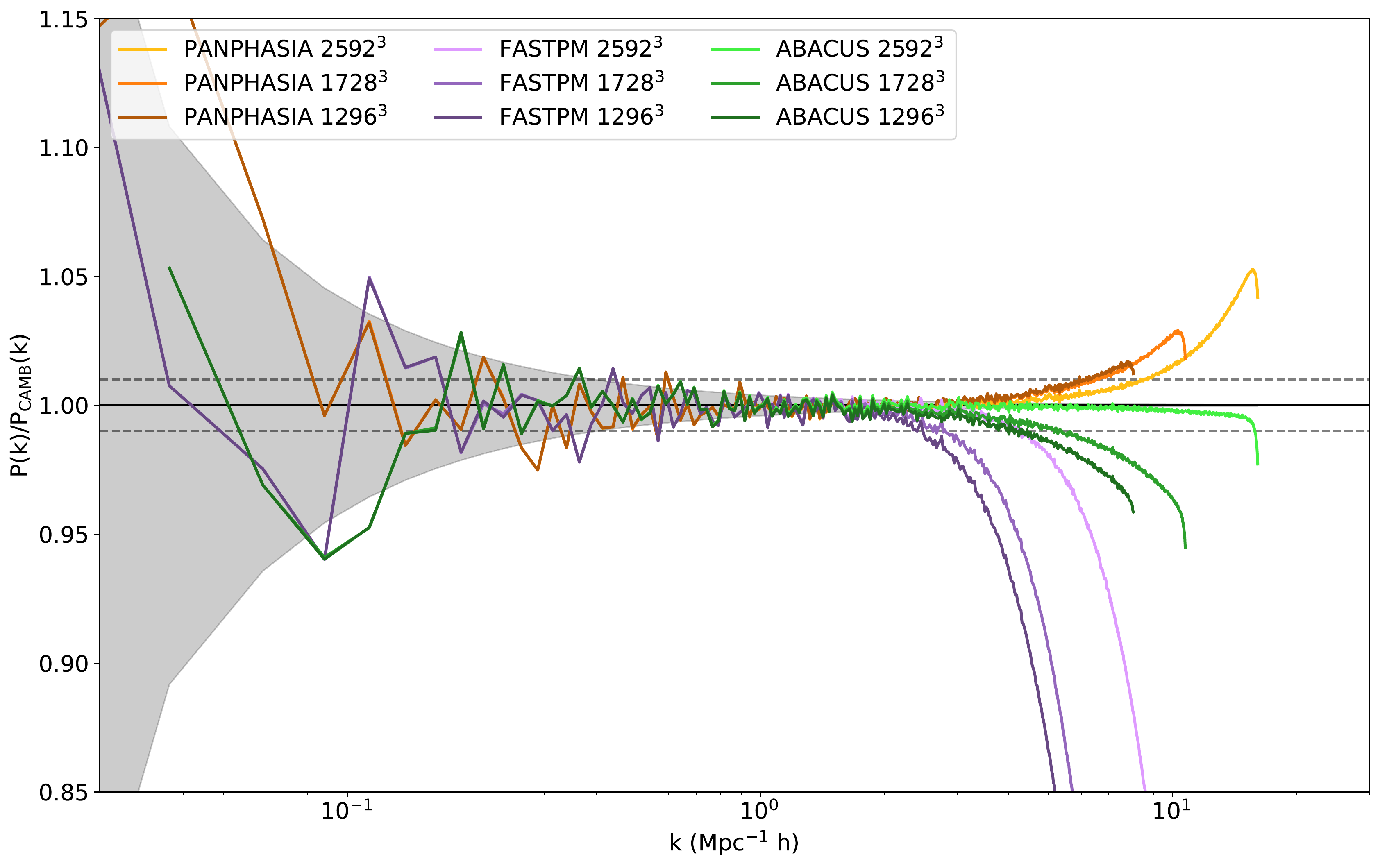}
    \caption{A comparison of the power spectrum of initial conditions created at redshift $z = 199$, using different codes and resolutions.  The power spectra are plotted as a ratio to the input theoretical power spectrum from \textsc{CAMB}.  At high $k$ differences emerge between the input and the measured power spectra.  At low $k$, the sample variance causes noise in the measured power spectrum.  The grey envelope shows the theory 1-sigma sample variance.The dashed lines indicate $\pm1$\%.}
    
    % Add to caption the low resolution curves are under the high resolution ones for abacus, panphasia
    \label{fig:IC_compare}
\end{figure*}

Fig.~\ref{fig:IC_compare} shows the power spectra measured for all ICs codes and also for different resolutions, compared to the reference power spectrum. The reference power spectrum is the theoretical linear power spectrum from CAMB at redshift, $z =199$. The measured power spectra differ from the reference power spectrum at both large (low $k$) and small scales (high $k$). The large scale differences are mainly due to sample variance and the limited number of large scale modes in a particular realisation may lead to disagreement in the  power spectrum measurement from the reference $P_{k}$ at these scales. The grey envelope shows the theory sample variance.  This variance decreases as $k^{-1}$ which is consistent with the theory expectation using Poisson noise where the number of modes is proportional to $k^{2}$.

There is only one distinct line for each code at low $k$, this is because each code uses a different phase realization which is consistent between different resolutions. At small scales, close to the Nyquist frequency, the systematic resolution effects are different between codes. The ICs created using \textsc{panphasia} do not disagree with the reference power spectrum by more than 5\% in all resolutions up to the Nyquist frequency.  The \textsc{fastpm} ICs have much lower power than the reference power spectrum for $k >  k_{\rm Nyquist}/2$.  The \textsc{abacus} ICs have the best agreement with the reference power spectrum.

It should be noted that it is not imperative for initial conditions to exactly fit the reference power spectrum at the smallest scales in order to create a realistic $N$-body simulation. The majority of the power at low redshift at small scales is due to collapse of larger modes and therefore small differences in the power spectrum around $k_{\rm Nyquist}$ are mostly unimportant to the growth of structure at these scales. 

%\subsection{Velocity statistics \& multiple realisations}
\subsubsection{Comparison of Velocity statistics}
\label{sec:ICstats}

The power spectrum is only a measure of the displacements of the particles produced by an initial conditions code. The particle velocities are just as important as the displacements and affect the rate of growth of structure as the simulation progresses. In methods of generating the initial conditions, such as the Zel'dovich approximation \citep{Zeldovich1970} and 2LPT \citep{Crocce_2006}, the particle velocities are not free to vary but are fixed by the particle displacements. For example in the Zel'dovich approximation  a particle's velocity is in the same direction as, and proportional to, the displacement from the particle's gridpoint. For this comparison project we thought it prudent to check the consistency between particle velocities from different IC codes.  We investigated the pairwise velocity dispersion between the codes and also compared to 10 realisations of \textsc{panphasia} ICs. There was no theoretical estimate generated therefore we aimed to verify that there were similar results between the codes with the differences being consistent with the sample variation from the 10 \textsc{panphasia} realisations.

The pairwise velocity dispersions are calculated by taking the pairwise velocity between pairs of particles separated by a distance $r$ along with the component of the velocity parallel and perpendicular to the vector between the particles.

For particles with positions and velocities $(\mathbf{x_{1}},\mathbf{v_{1}})$ and $(\mathbf{x_{2}},\mathbf{v_{2}})$ the parallel and perpendicular components are
\begin{equation}
    v_{\parallel} = \frac{(\mathbf{x_{1}}-\mathbf{x_{2}})\cdot(\mathbf{v_{1}} - \mathbf{v_{2}})}{|\mathbf{x_{1}}-\mathbf{x_{2}}|}
\end{equation}
and
\begin{equation}
    v_{\perp} = \sqrt{|\mathbf{v_{1}} - \mathbf{v_{2}}|^{2} - v_{\parallel}^{2}}
\end{equation}
respectively.
This gives velocity dispersions in the parallel and perpendicular directions of
\begin{equation}
    \sigma_{\parallel}^{2}(r) = \langle v_{\parallel}^{2}(r) \rangle -\langle v_{\parallel}(r) \rangle^{2}
\end{equation}
and
\begin{equation}
\label{eq:perp}
    \sigma_{\perp}^{2}(r) = \langle v_{\perp}^{2}(r) \rangle -\langle v_{\perp}(r) \rangle^{2} .
\end{equation}

The isotropy of the universe implies that the second term of equation \ref{eq:perp} is equal to zero.

The velocity anisotropy parameter 
\begin{equation}
    \beta = 1 - \frac{\sigma_{\perp}^{2}}{2\sigma_{\parallel}^{2}}
\end{equation}
measures the relative size of the parallel and perpendicular velocity dispersions.

\begin{figure}
	\includegraphics[width=\columnwidth]{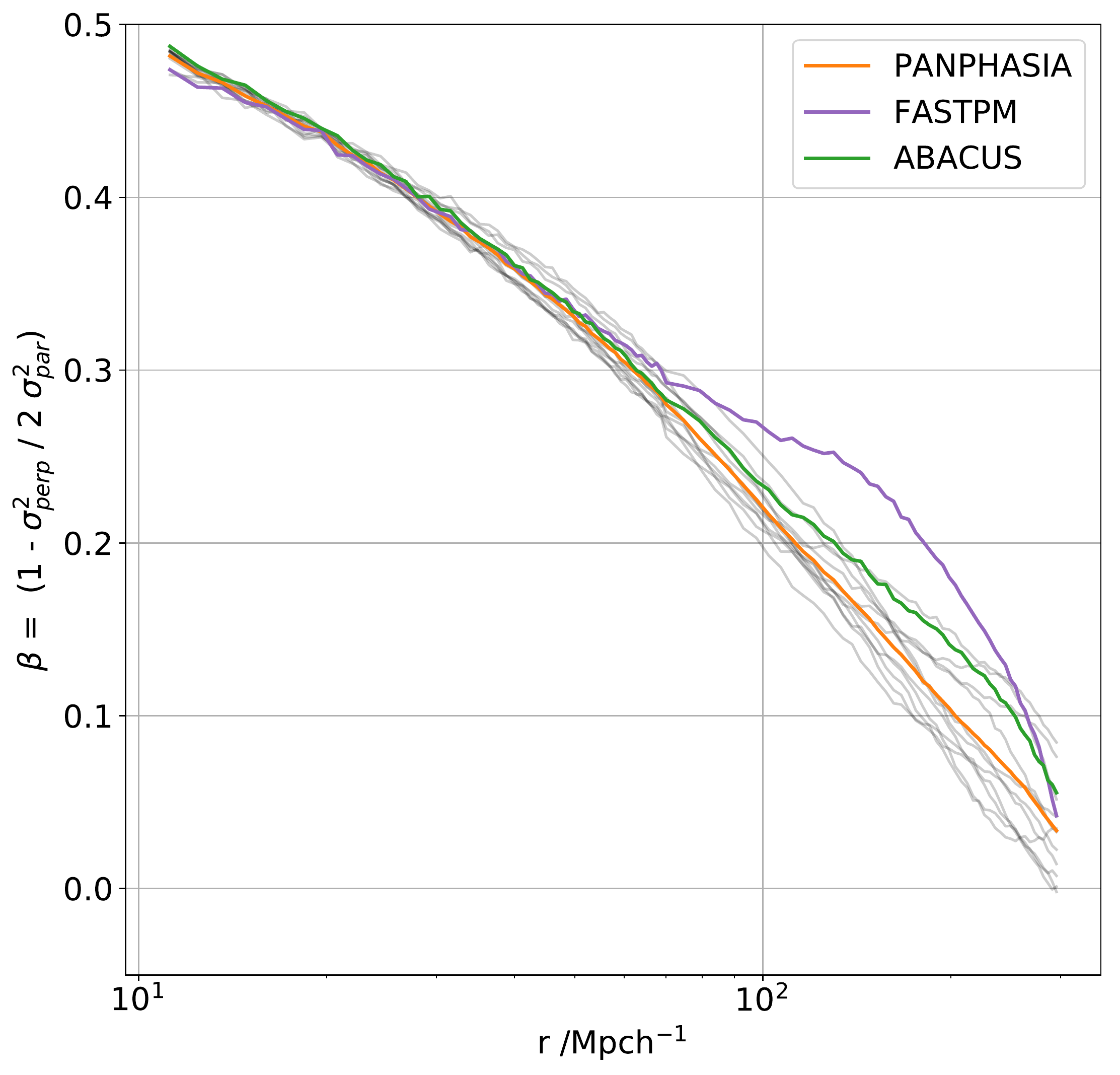}
    \caption{A plot showing the velocity anisotropy as a function of interparticle distance.  The coloured lines show the different codes meanwhile the grey lines show the results for 10 \textsc{panphasia} realisations to investigate sample variance.  The codes agree well at small distances while at large $r$ the differences are dominated by sample variance.  The only outlier is FastPM which has a higher anisotropy than any of the other realisations around 150~$h^{-1}$Mpc.}
    \label{fig:IC_velocity}
\end{figure}

Fig.~\ref{fig:IC_velocity} shows that the velocity anisotropy, as a function of pairwise inter-particle distance, is mostly consistent between the codes and at large scales the differences are mostly explained by sample variance.  There is excellent agreement between the codes at small inter-particle distance.  FastPM is the only outlier, having a slightly higher anisotropy than the other realisations at around $r = 150~h^{-1}$Mpc.

It was decided to use \textsc{panphasia} initial conditions for the $N$-body code comparison project in the next section. This was because \textsc{panphasia} had reasonable agreement with the reference power spectrum close to the Nyquist frequency in all resolutions and \textsc{panphasia} has the ability to easily share phase information to be used by other initial conditions codes in the future.

%\section{$N$-body Comparison}
\section{$N$-body Codes}
\label{sec:Nbody}

This section details the $N$-body code comparison project. We run \ABACUS, \SWIFT, \textsc{gadget-2}, and \FASTPM from $z=199$ to $z=0$ at the three resolutions shown in Table \ref{tab:1}. Comparisons between the codes are made in the matter power spectrum and in the dark matter halo properties, along with comparisons of the halo clustering.

The initial conditions for this code comparison project were created using \textsc{panphasia} and \textsc{ic\_gen} from the previous section; all codes used identical initial particles.  The cosmology used is described at the start of Section~\ref{sec:ICC} and the resolutions of the simulation boxes are shown in Table \ref{tab:1}.\footnote{The Panphasia descriptor describing the simulation volume is [Panph1,L21,(1133107,236124,673886),S81,CH1311607586,DESI\_IC\_v1]}.

Codes were run by separate groups and the code parameters were decided upon by those groups.

\subsection{Code details}
\subsubsection{\ABACUS}
%{\color{red} In each of these subsections there will be details on how each code works, any projects the code has been used in, the computing time and location of the runs etc. These can be submitted by those that ran the code}

\ABACUS is a high force accuracy $N$-body code that solves the far-field force with a high-order multipole method and the near field with direct summation on GPUs \citep{Abacus_code_paper}.  Typical RMS force accuracy is $10^{-5}$ to $10^{-6}$.  All particles in \ABACUS share a single, global time step, which is chosen at the beginning of each time step.  \ABACUS was recently used to run the AbacusSummit simulations \citep{AbacusSummit}, and the \ABACUS realization of the Euclid code comparison simulation was presented in \cite{Garrison2019}.

The numerical parameters used in this code comparison are the same as those used in AbacusSummit.  Specifically, spline force softening, fixed in proper coordinates, was used such that the $z=0$ value was 1/40th of the interparticle spacing (with the early time softening capped to 0.3 in the same units).  This choice was tested using scale-free simulations in \cite{2021MNRAS.504.3550G}.  The multipole order was $p=8$, with $405^3$ cells, and the time step parameter was set to $\eta_\mathrm{Accel} = 0.25$.  This is mildly more conservative than \ABACUS's run of the Euclid code comparison simulation, which used $\eta_\mathrm{Accel} = 0.3$.  \cite{Abacus_code_paper} show that a time step value of $\eta=0.25$ produces sub-percent shifts in the two-point clustering for all scales larger than twice the softening length.

\subsubsection{\GADGET}

\textsc{gadget-2}\citep{GADGET-2} combines a tree and a particle mesh for its gravity scheme. The first order moments are used to calculate the tree forces and a fixed opening angle criterion is used to decide when to divide up the tree cells.  The particle positions and velocities are updated in a leapfrog scheme and the global timesteps are evenly spaced in log($a$) with dynamic shorter timesteps for particles undergoing large accelerations. A mesh size of $N^{3}$ was used for the simulation with $N^{3}$ particles.  The comoving Plummer equivalent softening length used was 1/25th of the interparticle spacing. The parameters used to define the timestepping and force accuracy were MaxRMSDisplacementFac~$=0.2$, MaxSizeTimestep~$=0.01$, ErrTolTheta~$=0.5$, and ErrTolForceAcc~$=0.0025$.

\subsubsection{\SWIFT}

\SWIFT\footnote{\url{http://swift.dur.ac.uk/}}\citep{Schaller_2016} is a hydrodynamics and gravity code which uses a task based parallelisation strategy.  It is primarily designed for high resolution SPH simulations of galaxy formation, however \SWIFT can also be used for dark matter only simulations of large scale structure. The gravity scheme in \SWIFT has three levels. At large scales a Fourier mesh is used to calculate the force on particles and account for the periodicity of the simulation box; at small scales the fast multipole method \citep[FMM,][]{Dehnen2014} is used to calculate forces between particles in octree cells; at the smallest scales direct summation is used. A fixed opening angle criterion is used to calculate the size of the octree cells used in the FMM.  A time dependent force softening is applied to the particles, the forces are softened according to a spline kernel.  The global timesteps are evenly spaced in log($a$) and particles can have smaller timesteps if their acceleration exceeds a threshold \citep{Theuns2015}\footnote{\url{https://gitlab.cosma.dur.ac.uk/swift/swiftsim}}.  
A Fourier mesh size of 648$^{3}$ was used in the $1296^{3}$ particle and $2592^{3}$ particle simulations and $864^{3}$ was used in the $1728^{3}$ particle simulation.  These values were chosen as they were the largest allowed by the code which would not introduce aliasing with the initial particle grid.  The Plummer equivalent comoving softening length was 1/25th of the interparticle spacing.

\subsubsection{\FASTPM}
\FASTPM\footnote{\url{https://github.com/rainwoodman/fastpm}}\citep{FASTPM} is an approximated particle mesh $N$-body solver.  By modifying the kick and drift factors in time integration, it enforces correct linear displacement evolution on large scales, and meanwhile reduces the halo stochasticity when the number of time steps is small \citep{FASTPM}.  The cost is extremely low thanks to the small number of time steps required for satisfying statistics.  The accuracy can be mainly controlled by two options, $B$, the mesh size over particle number per dimension in the force calculation, and the time steps usually specified by the number of steps $T$ and the spacing scheme.  In this comparison project, we use $B=2$ and $T=46$, with linear spacing in the scale factor.
FastPM simulations are useful for estimating covariance matrices and other applications for which the accuracy is not a critical factor. A summary of the FastPM simulations prepared by the DESI Cosmological Simulations working group will be presented in \cite{Ding21}.

%\subsection{Power spectrum code comparison}
\subsection{Comparison of particle distribution}
\label{sec:PS}
\begin{figure}
	\includegraphics[width=\columnwidth]{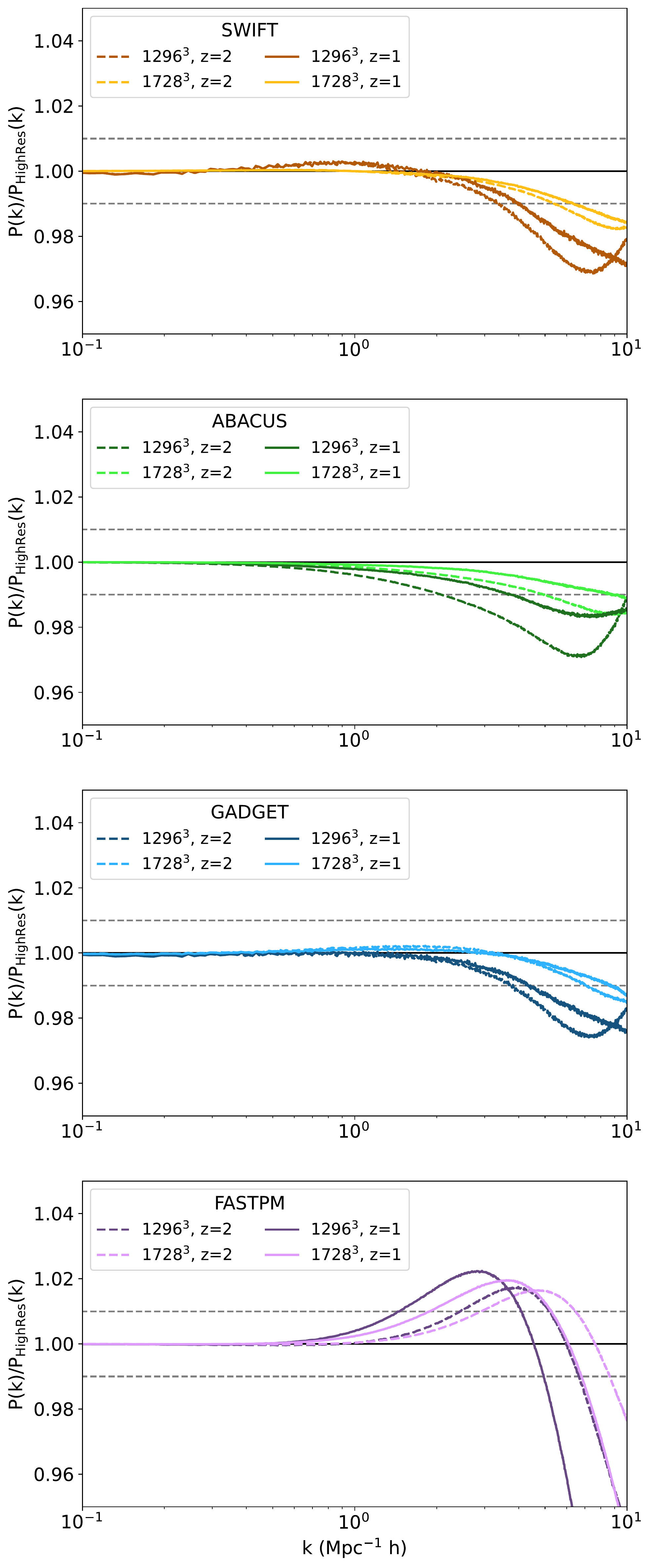}
    \caption{The matter power spectrum of the low and medium resolution runs relative to that of the high resolution run for each code at $z=1$ and $2$. \SWIFT, \ABACUS, \& \GADGET show similar behaviour where the power spectrum decreases by up to 2\% at $k=10~h$ Mpc$^{-1}$ when decreasing the resolution from $2592^3$ to $1728^3$ particles or from $1728^3$ to $1296^3$ particles. The lower resolution \FASTPM power spectra show an increase relative to the high resolution case at $k=3~h$ Mpc$^{-1}$ before decreasing sharply above $k=5~h$ Mpc$^{-1}$.}
    \label{fig:nbody_res_each_code}
\end{figure}

We first begin by examining the power spectrum dependence on the mass resolution for each code separately.  As seen in Fig~\ref{fig:nbody_res_each_code}, this dependence is similar for all the codes tested.  There is very little change in the power spectra at large scales and at small scales there are differences of 1-2\%.  The higher the mass resolution, the higher the power at small scales (except with \FASTPM).

\begin{figure*}
	\includegraphics[width=\textwidth]{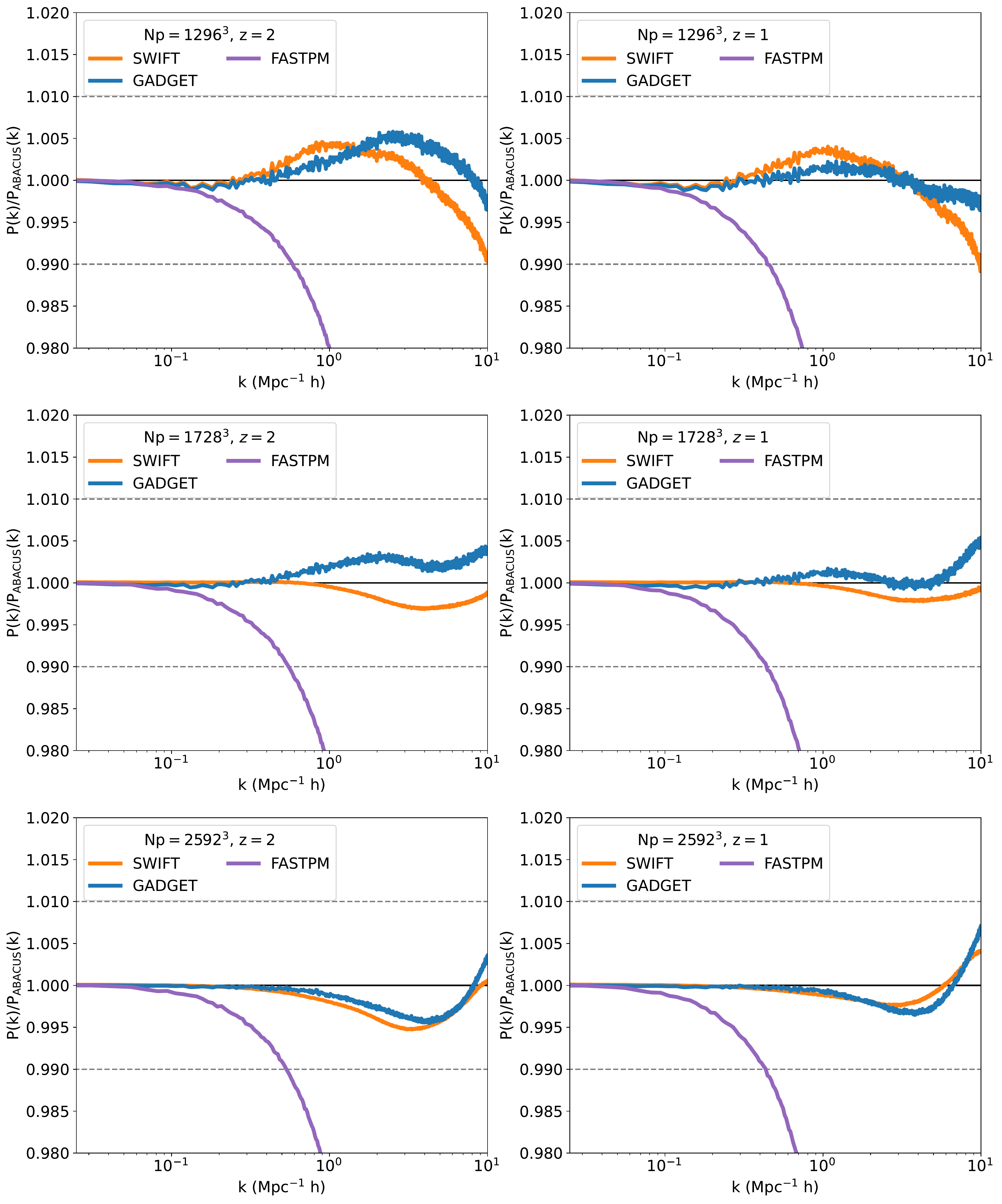}
    \caption{A comparison of the matter power spectrum between the different codes at $z=1$ and $z=2$ in all three resolutions. The power spectrum is plotted relative to \ABACUS at each redshift and resolution.  All the codes agree to within 0.1\% at $k<0.1~h$ Mpc$^{-1}$.  Differences emerge at high $k$ but, excepting \FASTPM, these are within 1\% at $k=10\, h$Mpc$^{-1}$.}
    \label{fig:nbody_code_each_res}
\end{figure*}

We then proceed to compare the power spectrum results among codes.  Fig.~\ref{fig:nbody_code_each_res} shows all of the power spectra relative to the \ABACUS runs.  There is broadly good agreement between all of the codes except \FASTPM with differences below 1\% at $k=2~h$ Mpc$^{-1}$. This agrees with the results found in other code comparison projects \citep{Schneider2016,Garrison2019}.  We do not see the discrepancy between \GADGET and linear theory at large scales as seen in \citep{Schneider2016}, although this work employs \textsc{Gadget-2} and \citeauthor{Schneider2016} employed \textsc{Gadget-3}.  Apart from \FASTPM, at $k=10~h$ Mpc$^{-1}$ the fractional differences between the power spectra are below 1\%.

\begin{figure}
	\includegraphics[width=\columnwidth]{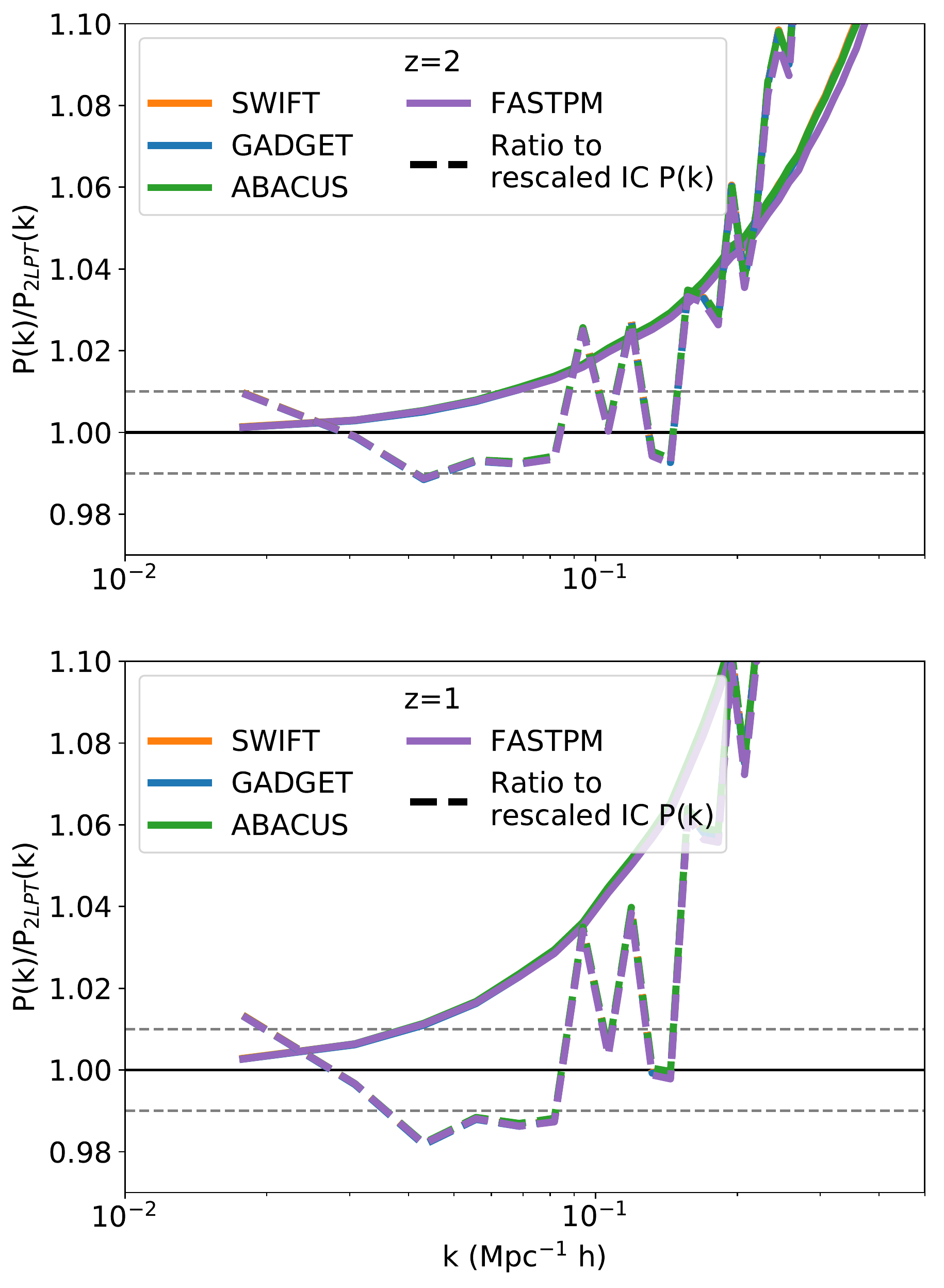}
    \caption{A comparison of the power spectra to perturbation theory at $z=1$ and $z=2$.  The perturbation theory power spectrum was found using second order perturbation theory (2LPT). The power spectra should asymptotically agree with 2LPT at low $k$. Coloured dashed lines show the ratio of the power spectra with initial conditions rescaled by the growth factor which do not agree as closely as 2LPT.}
    \label{fig:perturbation}
\end{figure}

All codes agree on the low-$k$ power spectrum amplitude to within 0.1\%, but to ascertain that this value is that predicted by perturbation theory, the 2LPT estimates of the $z=1$ and $z=2$ power spectra were measured by using \textsc{ic\_gen}, the code used to create the initial conditions for the simulations. Using a low output redshift for the same phases gives us the estimate of the matter distribution at low redshift from which the power spectrum can be measured. The results of this test are shown in Figure \ref{fig:perturbation}. We also attempted to compare to linear theory by rescaling the power spectrum of the initial conditions by the linear growth factor, however this did not reproduce the large scale power spectrum to sub-percent accuracy, as shown in the dashed lines in Figure \ref{fig:perturbation}. The reason for this difference is likely to be mode coupling.  In addition, using the Zel'dovich approximation instead of 2LPT produced a curve which was less smooth and therefore we used 2LPT for the low redshift perturbation theory power spectrum. Perturbation theory is only accurate for small density perturbations which corresponds to very large scales at low redshift. We have found that the simulation power spectra and the 2LPT power spectra asymptotically converge at low $k$ for all the codes.  Furthermore, this exercise helped identify a previously unknown error in the large-scale growth in the \SWIFT code; a difference of around 0.5\% in the large scale power spectrum which was caused by incorrect force integration. The error was corrected, and all results in this work use the modified code.

\section{Dark Matter Halo Comparison}
\label{sec:Halo}
%This section shall include the comparison of halo properties between different resolutions and different codes.

Much of the value which comes from running large $N$-body simulations does not come from the complete distribution of dark matter particles, but instead comes from grouping the particles into overdense regions called dark matter halos because these are the sites that host galaxies \citep{Peacock2000,Benson2000,Conroy2009}. Using and storing full particle data can become prohibitively expensive for large $N$-body simulations therefore halo catalogs are often the most useful data product to be produced from an $N$-body simulation. Observations and simulations suggest galaxies reside within dark matter halos and therefore they are essential to connect $N$-body simulations to observable results \citep{Wechsler18}.

 Halo catalogs can be used to create mock galaxy catalogs which provide useful information on the expected observational results for different cosmologies, along with providing mock data which is essential for testing analysis pipelines, e.g. see \cite{cole1998,Rodriguez-Torres:2015vqa,derose2021dark}.

%------Chani is adding---------------
In this section we perform the comparison of halo properties, mass functions and clustering. Halos are identified with the help of the phase–space temporal halo finder \ROCKSTAR with force resolution parameters shown in Table~\ref{tab:rockstar} \citep{Behroozi2013}\footnote{https://bitbucket.org/gfcstanford/rockstar}, and default parameters otherwise, including \texttt{STRICT\_SO\_MASSES=0}.

\begin{table}
\caption{\ROCKSTAR force resolution: this table shows the force resolution of the simulations in comoving $h^{-1}$Mpc, as input to the \ROCKSTAR halo finder. Halos whose centres are closer than the force resolution are considered to be unresolved. The minimum number of particles considered to be
a halo seed for $1296^3$, $1728^3$ and $2592^3$ resolutions are $5, 10$ and $10$ respectively.}
	\centering
	\begin{tabular}{lllcr} 
		\hline
		    Simulation & \ABACUS &  \GADGET & \SWIFT \\
		\hline
		$1296^3$ & 0.015  & 0.015  & 0.015  \\
		$1728^3$ & 0.0072  & 0.011  & 0.011      \\
		$2592^3$ & 0.0048 & 0.0077  & 0.0077          \\
			\hline
	\end{tabular}
	\label{tab:rockstar}
\end{table}

We do not include \FASTPM in this comparison of halo properties and clustering. This is because performing abundance matching is outside the scope of this paper, and the \FASTPM results without abundance matching would be unrealistically discrepant compared with how the code is used in practice.

Our tests involving dark matter halos are performed at $z=1$. We do not observe significant redshift evolution in our code or resolution comparisons when measuring halo mass functions and halo clustering at $z=0$ and $z=2$.

\subsection{Halo properties} 
\label{sec:Haloprop}

\begin{figure}
	\includegraphics[width=\columnwidth]{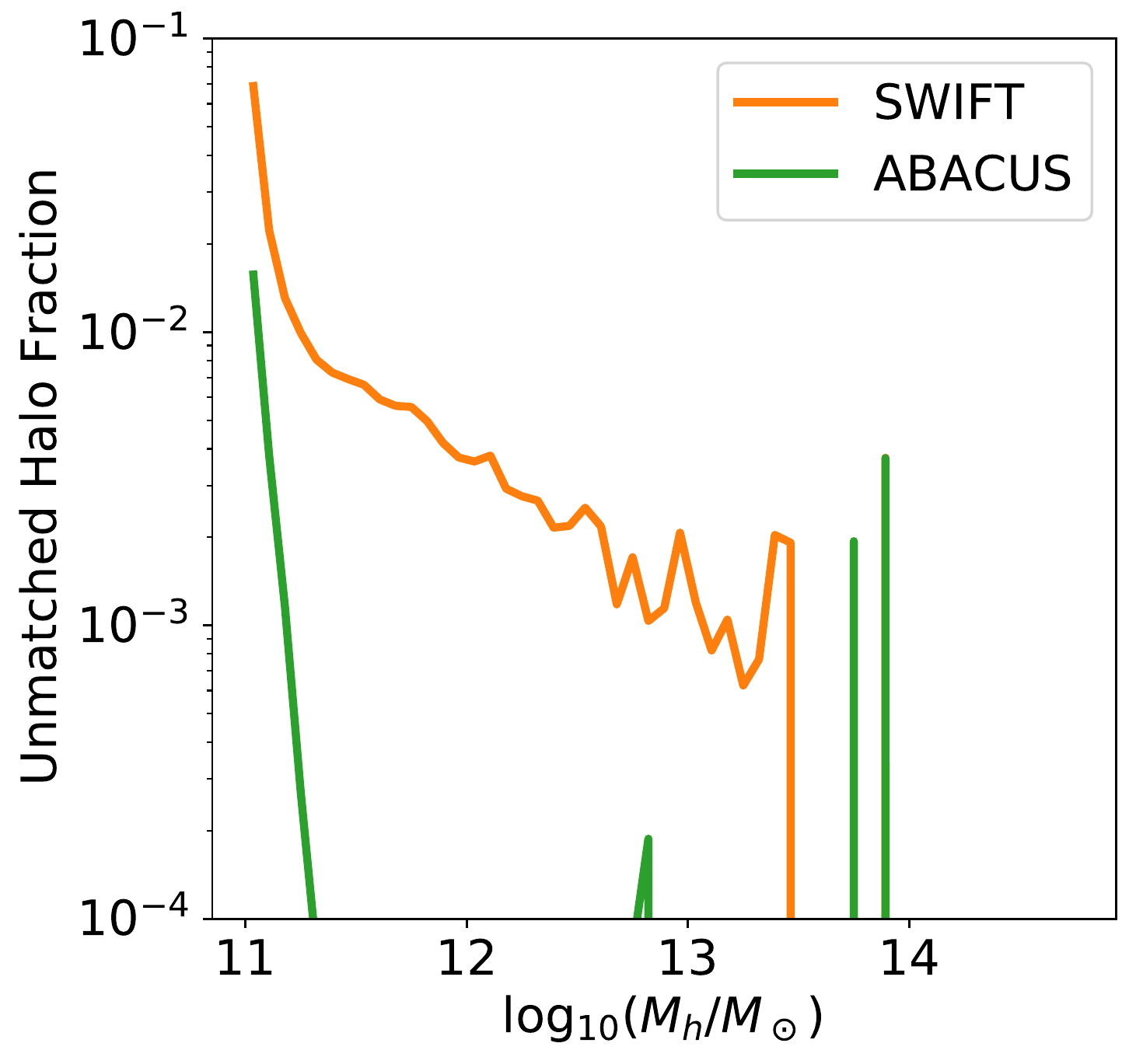}
    \caption{The fraction of halos which remain unmatched as a function of halo mass in the \ABACUS and \SWIFT catalogs when matching halos between the high resolution catalogs. The majority of halos are matched across simulations.}
    \label{fig:missing_halos}
\end{figure}

\begin{figure*}
	\includegraphics[width=\textwidth]{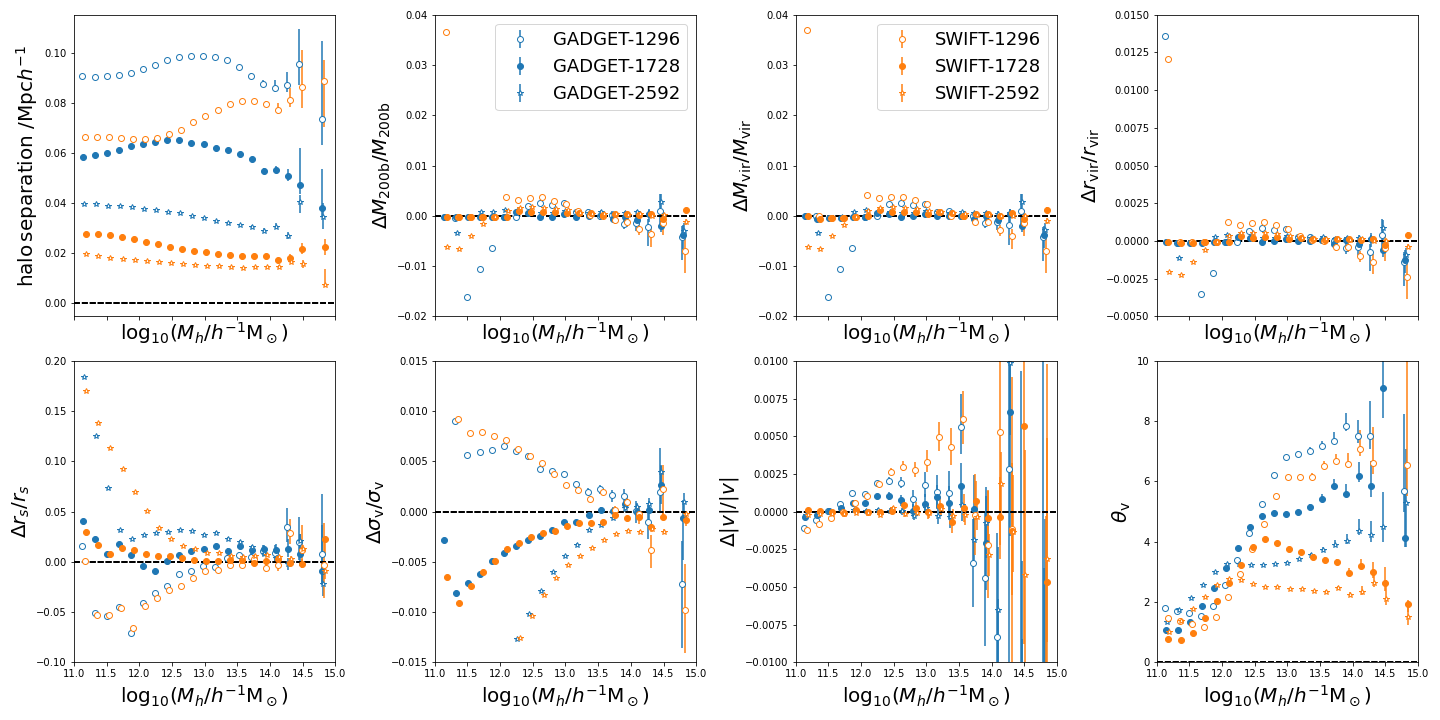}
    \caption{A comparison of the halo properties for halos matched between \ABACUS and the other codes at fixed resolution. On the top row from left to right the panels show: 1) The distance between matched halo centres in $h^{-1}$Mpc 2) The fractional difference in halo mass, as measured within a sphere of density 200 times the background density. 3 \& 4) The fractional difference in the virial mass and radius of matched halos.
    On the lower row, from left to right, the panels show: 1) the fractional difference in the scale radius of matched halos. 2) The fractional difference in the velocity dispersion of matched halos. 3) The fractional difference in the magnitude of the velocities of the matched halos. 4) The difference in the direction of the velocities of matched halos in degrees. Hollow circles, filled circles, and stars represent the comparison for the $1296^{3}$, $1728^{3}$, and $2592^{3}$ simulations respectively. The results are colour coded differently for each simulation code. \GADGET: blue, \SWIFT: orange, and \ABACUS is used as the reference simulation.}
    \label{fig:halo_prop_code}
\end{figure*}

\begin{figure*}
	\includegraphics[width=\textwidth]{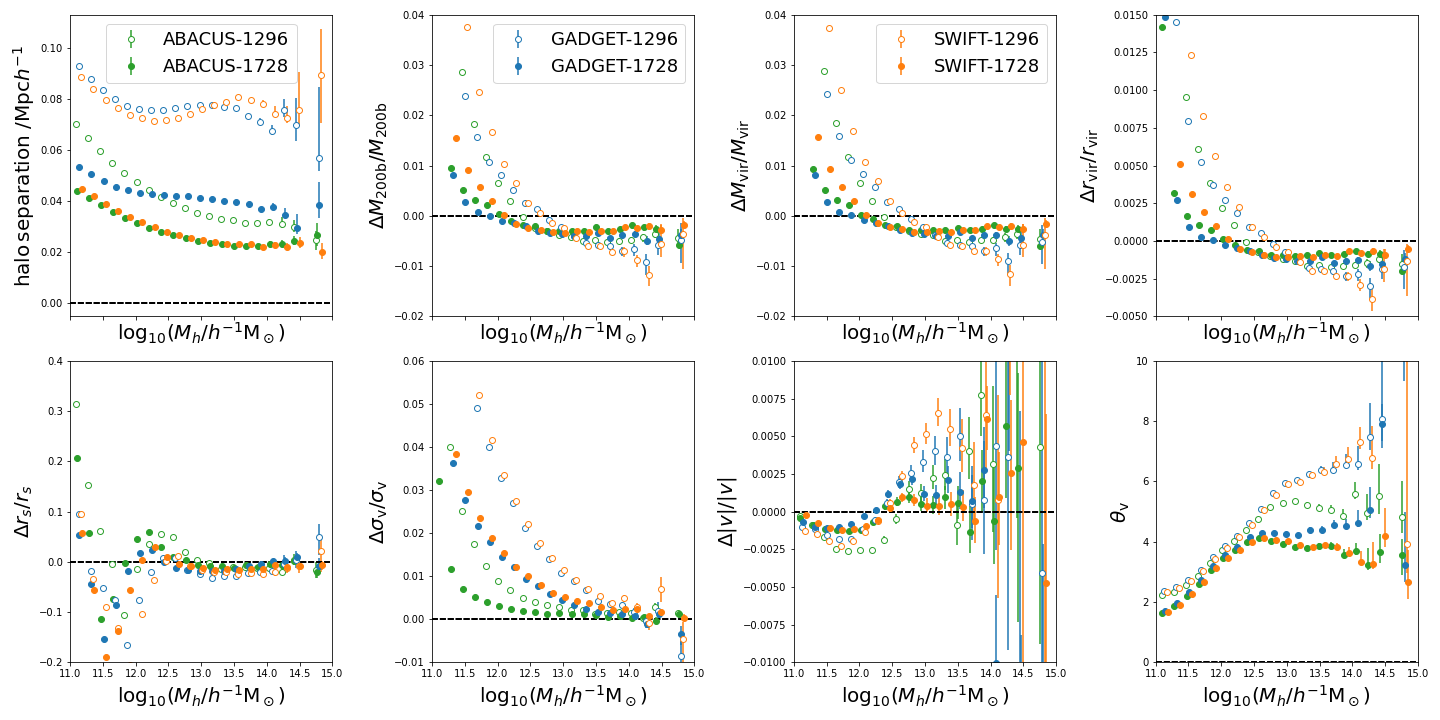}
    \caption{As for Fig.~\ref{fig:halo_prop_code} but for halos matched to the high resolution simulation for each code. Note the change in range plotted in the bottom left panels.}
    \label{fig:halo_prop_resolution}
\end{figure*}

We execute a detailed halo-to-halo comparison by matching halos between $N$-body simulations and comparing their properties.  There is no guarantee that halos will form in identical places, but since all simulations share the same initial particles, in many cases a robust match can be identified. Still, matching halos provides a challenge as there is no unique identifier for each halo which is well defined across several different simulation runs.

We use a nearest neighbour approach to match halos between snapshots from simulations run with different codes and resolutions. Each halo in one catalog was matched to the halo with the nearest position in the other catalog. All halos with a mass smaller than $5 \times 10^{10}~$M$_{\odot}/h$ at all resolutions were removed to avoid spurious matches and an upper distance limit of $0.25\, h^{-1}$Mpc was applied for halos to be considered as matched. The matched halo properties were binned by halo mass.

Figure~\ref{fig:missing_halos} shows the fraction of unmatched halos in each catalog in a comparison between the high resolution \SWIFT and \ABACUS simulations. This fraction increases as halo mass decreases but is only greater than 5\% of halos for the lowest mass \SWIFT halos at $M_{h}=10^{11}$~$h^{-1}$M$_{\odot}$. This gives us confidence that there are not systematic effects being introduced by the nature of the halo matching algorithm.

The halo properties which are compared between matched catalogs are defined as follows. Halo separation is the distance between matched halos in units of $h^{-1}$Mpc, as discussed above this cannot be greater than  $0.25 \,h^{-1}$Mpc for any matched pair. $M_{\textrm 200b}$ is defined as the mass contained within a sphere around the halo such that the sphere has a density of 200 times the background density of the universe. $r_{\rm vir}$ is the radius of the halo which is similarly defined in terms of a spherical overdensity. In this case the overdensity is $\Delta_{c}$ times the critical density of the universe.  $\Delta_{c}$ is the solution for a virialised cluster which is $18\pi^{2}$ for a critical universe but can vary with redshift otherwise. $M_{\rm vir}$ is the mass contained within $r_{\rm vir}$ \citep{Peebles1980,Eke1996,Bryan1998}. $r_s$ is the NFW scale radius of the halo, defined in terms of the density profile \citep{Navarro1997}. $\sigma_{\rm v}$ is the halo particle velocity dispersion in physical km~s$^{-1}$. $|v|$ is the speed of the halo in physical km~s$^{-1}$. Finally $\theta_{\rm v}$ is the difference in angle of the velocity vectors of the matched halos in degrees.  With the exception of halo separation, $|v|$, and $\theta_{\rm v}$, which we compute ourselves, all of these properties are computed by \ROCKSTAR.

Fig.~\ref{fig:halo_prop_code} shows how the matched halo properties vary between \textsc{abacus} and the other codes at fixed resolution. The low, medium, and high resolution cases are shown as hollow circles, solid circles and stars respectively along with error bars which show the error in the mean of the matched property within each mass bin.

In general, the matched properties show greater agreement at high masses and at high resolution. This is where the halos contain a greater number of particles and therefore it is to be expected that these halos are less susceptible to variation due to differences between the codes which are primarily in the small scale forces. The halo separation is roughly constant across halo mass but varies for different codes and resolutions. The mean halo separation varies between 10 and 100~$h^{-1}$kpc; \SWIFT, and \ABACUS agree to within three softening lengths at high and medium resolution, while \GADGET has two to three times worse agreement.

\textsc{swift} and \textsc{gadget} show agreement with \textsc{abacus} to within 1\% in the halo mass and radius at medium and high resolution for all masses. At low resolution differences emerge for halo masses below $10^{12}$~$h^{-1}$M$_{\odot}$.

$r_s$ shows a similar pattern of better agreement at higher halo masses. The differences at masses below $10^{12}$ $h^{-1}$M$_{\odot}$ are at the 10\% level and do not show clear trends with code or resolution, except that \textsc{swift} and \textsc{gadget} show progressively lower $\Delta r_s$ at lower resolutions.

As resolution decreases, \SWIFT and \GADGET have increasing $\Delta \sigma_{\rm v}$ relative to \ABACUS.

The halo velocities show larger differences in both magnitude and direction between the codes for high mass halos. There are large uncertainties on these measurements due to the limited number of high mass halos within each catalog.

Fig.~\ref{fig:halo_prop_resolution} shows how the matched halo parameters vary with simulation resolution when keeping the simulation code fixed.

There is a clearer trend in the halo separation with halo mass when comparing with each code fixed, the separation is lower for higher mass halos and is also lower at higher resolution. \SWIFT and \GADGET show larger halo separation at different resolutions compared to \ABACUS.

The halo masses are systematically smaller by around 0.5\% for all codes and resolutions at halo masses above $10^{13}$~$h^{-1}$M$_{\odot}$. The same feature is seen in the virial radii where the difference is 0.2\%. For lower mass halos the opposite effect is seen, halo masses are larger for the lower resolution simulations and this difference grows when going to lower halo masses and lower resolutions. This indicates that at lower resolutions mass is systematically moved from high mass halos into low mass halos when compared to higher resolution simulations. The origin of this effect could be that the halo finder is more likely to split off halos at low resolutions due to a lower number of particles with which to make friends-of-friends connections.

$r_{s}$ is consistent between high mass halos at all resolutions for all the codes. Below $10^{12}$ $h^{-1}$M$_{\odot}$, $r_{s}$ becomes poorly matched as resolution is varied.

The velocity dispersion, $\sigma_{\rm v}$, increases for lower mass halos at lower resolutions. This effect can be seen for all the codes but is most strongly present in \SWIFT, and \GADGET.

In summary, the halo properties agree well between the different codes at fixed resolution, there is better agreement for halos with mass greater than $10^{13}$~$h^{-1}$M$_{\odot}$ than lower mass halos. 

Lower simulation resolution systematically biases high mass halos to have slightly lower mass and radius than those matched from higher resolution simulations. The opposite effect is observed at low masses. In general high mass halos show greater agreement in the matched halo parameters than low mass halos.

\subsection{ Halo Mass Function}
\label{sec:HMF}

The halo mass function (HMF) describes the number density of dark matter halos in different mass intervals. Here we have defined the HMF as ${\textrm{d}n}/{\textrm{d}\textrm{log}_{10}(M_{\textrm{vir}})}$ where $n$ is the number density of halos in units of $\textrm{Mpc}^{-3}h^{3}$ and $M_{\textrm{vir}}$ is in units of $h^{-1}$M$_{\odot}$.

The HMF has a strong effect on the number density and clustering of galaxy catalogs created using the halo catalog, particularly at fixed mass. It is therefore important for the HMF to not change significantly between simulations run with different codes, as this could indicate underlying systematic errors.

In addition we investigate the effect of simulation resolution on the HMF.  Mass resolution has the greatest effect at the low mass end of the HMF where halos contain fewer particles.  It is important to find the cutoff mass below which the HMF shows significant differences due to simulation resolution.

\begin{figure}
	\includegraphics[width=\columnwidth]{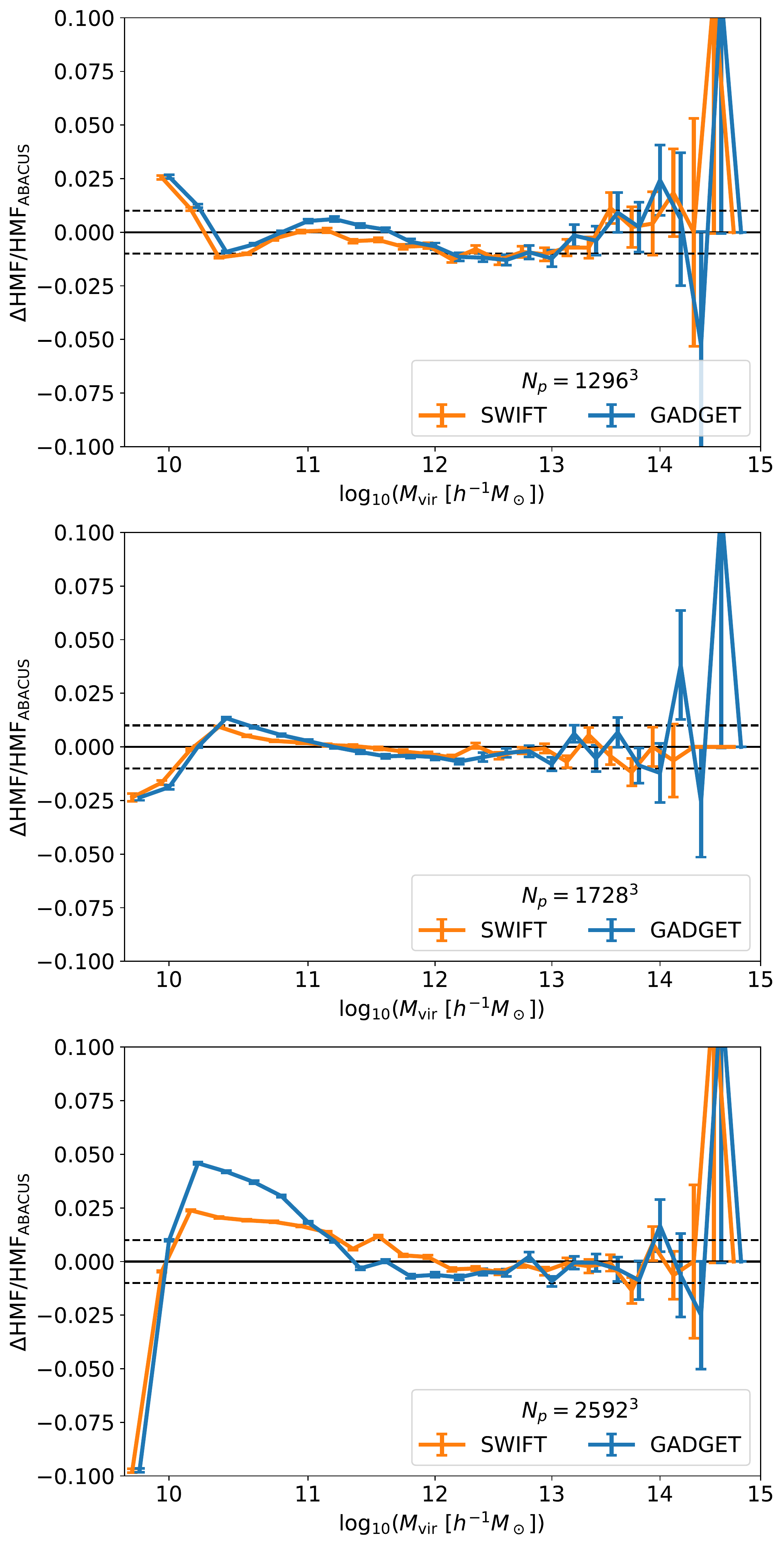}
    \caption{The ratio of the halo mass functions to that of the \ABACUS simulation at fixed resolution at $z=1$. The top, middle and bottom panels show the low, medium, and high resolution simulations respectively. The errorbars show an estimate for the noise while comparing two simulations which share the same phases. Differences between the codes are below 1\% between $10^{12}~h^{-1}$M$_{\odot}$ and $10^{13.5}~h^{-1}$ M$_{\odot}$ at all resolutions. The differences in the HMF at large halo masses are consistent with noise. At low halo masses there is better agreement between the codes at low resolution than high resolution, in the high resolution comparison the HMFs differ by up to 5\% between $10^{10}~h^{-1}$M$_{\odot}$ and $10^{11}~h^{-1}$ M$_{\odot}$ meanwhile in the medium and low resolutions the differences do not exceed 2.5\% across the same mass range.}
    \label{fig:HMF_panel}
\end{figure}

Figure \ref{fig:HMF_panel} shows the HMFs measured from the different codes at fixed resolution at $z=1$, relative to the \ABACUS HMF. The errorbars represent the estimated standard deviation in the HMF difference between phase matched simulations which are calculated using a jackknife method, see equation \ref{eq:jk_noise} in Section \ref{sec:Halo_Clustering} for more details. Differences between the codes are greater at high and low masses. The HMFs agree at the 1\% level between $10^{12}~h^{-1}$M$_{\odot}$ and $10^{13.5}~h^{-1}$ M$_{\odot}$ for all resolutions.  The HMFs do not differ by greater than 10\% at masses up to $10^{14.5}~h^{-1}$ M$_{\odot}$, these differences are consistent with noise and not biased in a particular direction.  At low masses there is different behaviour between the codes at each resolution. There is better agreement on the low mass end of the HMF at low resolution than high resolution.

\begin{figure}
	\includegraphics[width=\columnwidth]{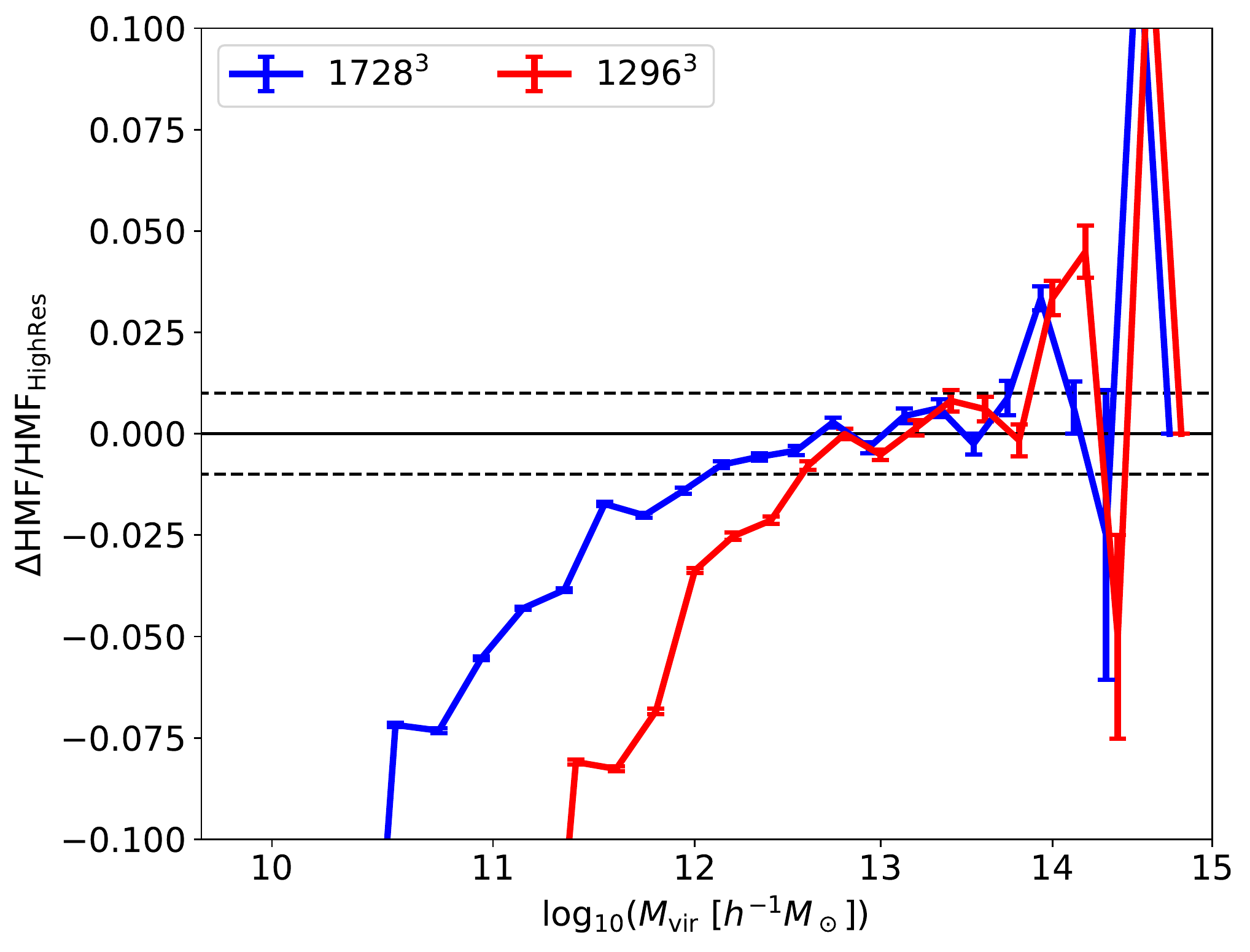}
    \caption{The ratio of the halo mass functions from the \ABACUS simulations relative to the high resolution run at $z=1$. The errorbars show an estimate for the noise while comparing two simulations which share the same phases. The $1296^{3}$ simulation shows greater differences than $1728^{3}$ to the $2592^{3}$ simulation. There is a deficit in low mass halos in the lower resolution simulations relative to the high resolution simulation.}
    \label{fig:HMF4}
\end{figure}

Fig.~\ref{fig:HMF4} shows the change in HMF with resolution for the \ABACUS code at $z=1$. The effect of changing resolution is larger than changing code. The halo catalogs from different resolution simulations have similar HMFs at for high mass halos. The medium resolution simulation HMF differs from the high resolution simulation HMF by more than 1\% below $10^{12}~h^{-1}$ M$_{\odot}$. The low resolution HMF differs by a greater amount than the medium resolution HMF, with greater than 1\% disagreement below $10^{12.5}~h^{-1}$ M$_{\odot}$. 

%\subsection{Halo Clustering}
\subsection{Halo Clustering: Correlation Function}
\label{sec:Halo_Clustering}

The clustering of dark matter halos is one of the most important statistics to be produced in an $N$-body simulation. Galaxy clustering is closely tied to halo clustering, especially at large scales, therefore ensuring accurate and robust halo clustering estimates is essential in order to have confidence in mock data produced from simulations.

In this section we compare the halo clustering from simulations run with different codes and at different resolutions. We compare the clustering in both real-space and in multipoles of redshift space. Estimates of the DESI statistical error are made using a jackknife method, allowing us to provide length scales above which the halo clustering from our simulations is robust for DESI analysis.

The two-point correlation function for halos, $\xi(r)$ is defined as the excess probability of finding a halo at a distance $r$ from another halo, averaged over all halos.

\begin{equation}
    \xi(\abs{\vb{r}}) = \langle \delta(\vb{x})\delta(\vb{x}-\vb{r}) \rangle
\end{equation}

\begin{figure}
	\includegraphics[width=\columnwidth]{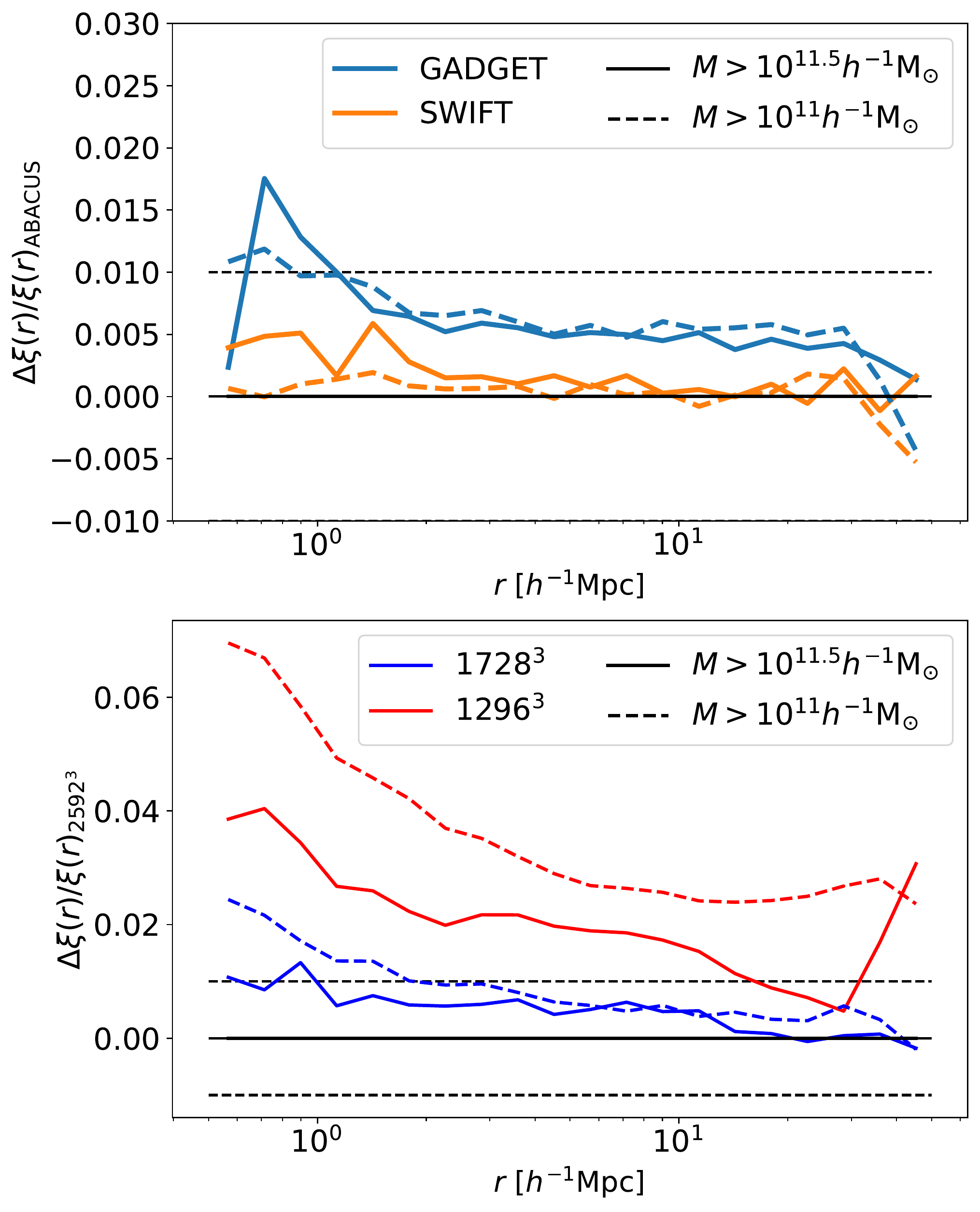}
    \caption{Upper Panel: Real-space 2-point correlation functions of the halos from the 2592$^{3}$ simulations at $z=1$. The results from using two halo mass limits are shown as solid and dashed lines. \SWIFT and \ABACUS agree more closely than \GADGET and \ABACUS but for $r>1$Mpc all the codes agree to within 1\%. There are not significant differences which come from changing the halo mass limit. Lower Panel: Real-space two-point correlation functions of the halos from the \ABACUS simulations relative to the high resolution simulation at $z=1$. Differences between the halo clustering measurements are larger at small scales. A higher halo mass limit and higher mass resolution both improve the agreement to the high resolution 2-point correlation function.}
    \label{fig:2PCF_both}
\end{figure}

%Where $r = |\vb{x}_1 - \vb{x}_2 |$
Fig.~\ref{fig:2PCF_both} shows the real-space two-point correlation functions of the halo catalogs from the different codes relative to \ABACUS. The high resolution simulations were used with a mass cut of $M > 10^{11.5}$~$h^{-1}$ M$_{\odot}$ and $M > 10^{11}$~$h^{-1}$ M$_{\odot}$ (solid and dashed lines respectively). The correlation functions from the different simulation codes do not differ by greater than 1\% for length scales above $1\, h^{-1}$Mpc.

In Fig.~\ref{fig:2PCF_both} we compare the real-space two-point correlation function of the halos from \ABACUS simulations at different resolutions. The high and medium resolution simulations agree to within 1\% for all length scales above $1\, h^{-1}$Mpc. A larger difference is seen in the low resolution halo clustering, the clustering increases at smaller scales to 4\% greater than the high resolution below $1\, h^{-1}$Mpc.

Decreasing the halo mass cut to $M > 10^{11}$~$h^{-1}$ M$_{\odot}$ can be seen to reduce the agreement in the clustering relative to the higher mass cut. This is seen most strongly the low resolution simulations.

The multipoles of the redshift space correlation function, $\xi_{l}(s)$ are defined   as 
\begin{equation}
    \xi_{l}(s) = \frac{2l + 1}{2}\int^{\pi}_{0}\textrm{d} \theta \sqrt{1 - \mu^{2}}\xi(\sigma,\pi)\mathcal{L}_{l}(\mu) 
\end{equation}
\citep{Peebles1980,Kaiser87}. Here $l$ is the multipole order, $\theta$ is the angle between the line of sight and the halo separation vector, $\mu = \rm{cos}(\theta)$, $\xi(\sigma,\pi)$ is the 2d correlation function, $\mathcal{L}_{l}(\mu)$ is the $l^{\rm{th}}$ Legendre polynomial.

When making these comparisons of halo clustering, it is useful to understand the tolerance of observational measurements as this will provide a target accuracy. We estimate this observational tolerance by estimating the statistical error in the clustering using the jackknife method \citep{Jackknife,Norberg2009}.

\begin{figure}
	\includegraphics[width=\columnwidth]{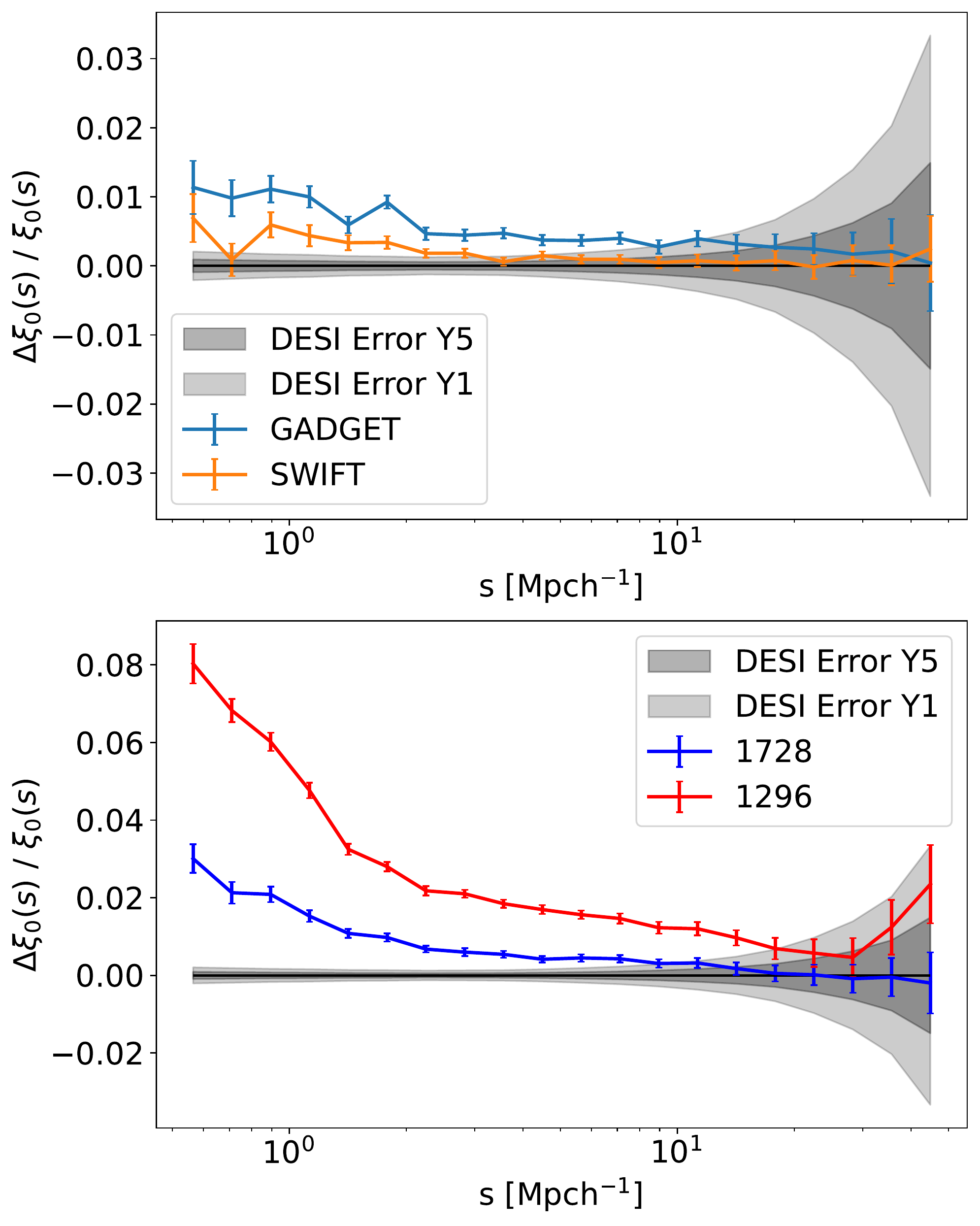}
    \caption{Upper panel: The redshift space clustering monopole comparison between the different codes at high resolution. \ABACUS is used as the reference clustering. The shaded regions show an estimate for the DESI survey year 1 and year 5 statistical errors. The errorbars show an estimate for the noise while comparing two simulations which share the same phases. A halo mass cut of $M > 10^{11.5}$~$h^{-1}$ M$_{\odot}$ was used. The systematic differences between the codes are within the expected year 5 DESI volume statistical errors for $s > 20\, h^{-1}$Mpc and within the year 1 volume statistical errors for $s > 10\, h^{-1}$Mpc. Lower panel: As above but comparing the redshift space clustering between the different resolutions for the \ABACUS code, with reference to the high resolution simulation. The medium resolution simulation is consistent with the high resolution simulation to within the DESI statistical errors for $s > 20\, h^{-1}$Mpc. The low resolution clustering is not consistent with the high resolution clustering to within the expected year 5 DESI statistical errors at any of the measured scales.}
    \label{fig:xil_0_both}
\end{figure}

\begin{figure}
	\includegraphics[width=\columnwidth]{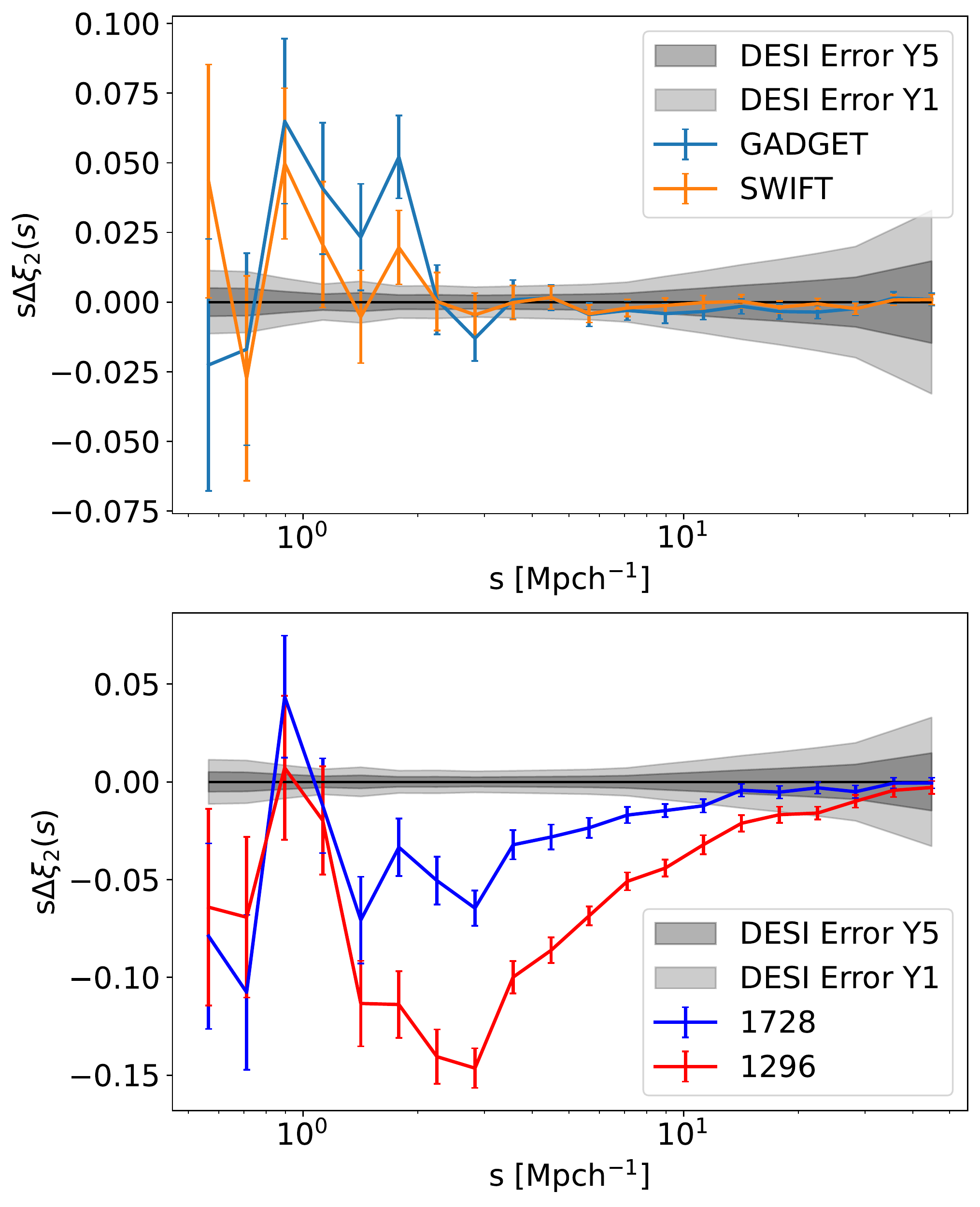}
    \caption{Upper panel: The redshift space clustering 2nd order multipole comparison between the different codes at high resolution. \ABACUS is used as the reference clustering. The shaded regions show an estimate for the DESI survey year 1 and year 5 statistical errors. The errorbars show an estimate for the noise while comparing two simulations which share the same phases. A halo mass cut of $M > 10^{11.5}$~$h^{-1}$ M$_{\odot}$ was used. The codes agree to within the DESI statistical errors for $s > 10\, h^{-1}$Mpc, below which noise begins to dominate the observed differences. Lower panel: As above but showing the redshift space clustering 2nd order multipole comparison between the different resolutions for the \ABACUS code, with reference to the high resolution simulation. There are significant differences compared to the level of the DESI statistical errors in both the low and medium resolution simulations for $s < 10\, h^{-1}$Mpc.}
    \label{fig:xil_2_both}
\end{figure}

The \ABACUS high resolution box was used for jackknife estimates. This box was split into $N = 100$ distinct sub-volumes. The redshift space correlation function multipoles were measured on $N$ sub-samples created by removing a single sub-volume each time to provide $N$ clustering estimates. The standard error of the clustering can then be found to be:

\begin{equation}
    \sigma(s) = \sqrt{\frac{N-1}{N}\sum_{i=0}^{N}{\left( \xi^{i}(s) - \overline{\xi}(s)\right)}^{2}}
\end{equation}
where $\xi^{i}(s)$ is the $i^{\rm{th}}$ clustering estimate, produced by removing the $i^{\rm{th}}$ sub-volume and $\bar{\xi}(s)$ is the mean correlation function,
\begin{equation}
    \overline{\xi}(s) = \sum^{N}_{i=0}\frac{\xi^{i}(s)}{N} 
\end{equation}
The standard error estimate, $\sigma(s)$, corresponds to the statistical error for a cubic simulation box with side length $500\, h^{-1}$Mpc. For the full DESI survey volume we use an estimated volume of $20 \,h^{-3}$Gpc$^{3}$ and for the year 1 DESI volume we use $4 \,h^{-3}$Gpc$^{3}$\citep{levi2013desi}. Our estimate of the statistical error on the DESI survey is produced by multiplying by the square root of the volume ratio between the DESI survey and the simulation box
\begin{equation}
    \sigma_{\rm{DESI}}(s) = \sqrt{\frac{V_{\rm{sim}}}{V_{\rm{DESI}}}}\sigma(s) 
\end{equation}
$\sigma_{\rm{DESI}}(s)$ is represented as a grey shaded region on Figs.~\ref{fig:xil_0_both}\textendash\ref{fig:xil_2_both}. 

A source of error in the clustering comparisons is that we are comparing the results from simulation boxes of finite size. The simulations were run from identical initial conditions, which removes sample variance. However our comparisons between different simulations with the same initial conditions still contain noise in the sense that the differences between these simulations contains a noise component that would change randomly if we changed the initial conditions and compared a second matched pair of simulations. This noise has the potential to obscure the measurement of systematic differences due to code and resolution. We use jackknife method to provide an estimate for this noise by using the variance in the clustering difference between matched sub-samples
\begin{equation}
    \sigma_{\rm{NOISE}}(s) = \sqrt{\frac{N-1}{N}\sum_{i=0}^{N}{\left( \Delta\xi^{i}(s) - \overline{\Delta\xi}(s)\right)}^{2}}
    \label{eq:jk_noise}
\end{equation}
where $\Delta\xi^{i}(s)$ is the difference in the clustering measurements from two simulations with the $i^{\rm{th}}$ sub-volume removed and $\overline{\Delta\xi}(s)$ is the mean clustering difference. A detailed investigation of the jackknife method estimating the noise between two simulations sharing the same phases will be presented in \cite{Zhang21}.

$\sigma_{\rm{NOISE}}(s)$ is represented as error bars on Figs.~\ref{fig:xil_0_both}\textendash\ref{fig:xil_2_both}. The clustering difference between \ABACUS and each code is used for the code comparison noise estimate, while the clustering difference to the high resolution \ABACUS simulations is used for the resolution comparison noise estimate. 

For the redshift space correlation function measurements, the results from three orthogonal lines of sight are averaged in order to reduce sample variance \citep{Smith2020}.

Figs.~\ref{fig:xil_0_both} and \ref{fig:xil_2_both} show the redshift space correlation function comparison between the high resolution simulations for different codes in the monopole and quadrupole respectively. Differences between the codes in the monopole are significant relative to the DESI statistical error below $10$ Mpc$h^{-1}$, with sub $0.1\%$ agreement required for consistency within the DESI statistical errors at these length scales. In the quadrupole the simulation noise dominates at small scales, making it difficult to establish systematic differences between the codes. The redshift space correlation function monopole and quadrupole difference between catalogs from different codes is smaller than the DESI statistical error above $20\, h^{-1}$Mpc. \ABACUS and \SWIFT have remarkable agreements down to $2\,h^{-1}$Mpc.

Figs.~\ref{fig:xil_0_both} and \ref{fig:xil_2_both} show the comparison between resolutions for the \ABACUS simulations. The low resolution correlation function monopole is not consistent with the high resolution case to within the DESI statistical error at any length scale. The medium resolution agrees with the high resolution correlation function to within the DESI and simulation errors for $s > 20\, h^{-1}$Mpc in the monopole and quadrupole.

Our results suggest that our medium and high resolution simulations have sufficient control over systematic errors in the redshift space correlation function for dark matter halos in the regime where $s > 20 \, h^{-1}$Mpc for halos with mass larger than $10^{11.5}$~$h^{-1}$ M$_{\odot}$.

%%% Redshift space multipoles with errorbars for resolution and code comparison

\subsection{Halo Clustering: Power Spectrum}
\label{sec:Halo_PS}

The dark matter halo power spectrum is another useful metric to perform comparisons between $N$-body simulation codes. The definition of the real-space power spectrum can be found in Section~\ref{sec:ICPS}. The multipole expansion of the redshift space power spectrum is used to understand the effects caused by peculiar velocities. The definition of the power spectrum multipole expansion is similar to the correlation function multipole expansion
\begin{equation}
    P_{l}(k) = (2l + 1)\int^{1}_{0}\textrm{d} \mu P(k,\mu)\mathcal{L}_{l}(\mu),
\end{equation}
where $l$ is the multipole order, $\mu$ is the cosine of the angle between the line of sight and the halo separation vector, $P(k,\mu)$ is the 2d, anisotropic power spectrum, $\mathcal{L}_l$ is the $l^{\rm{th}}$ Legendre polynomial.

\begin{figure}
	\includegraphics[width=\columnwidth]{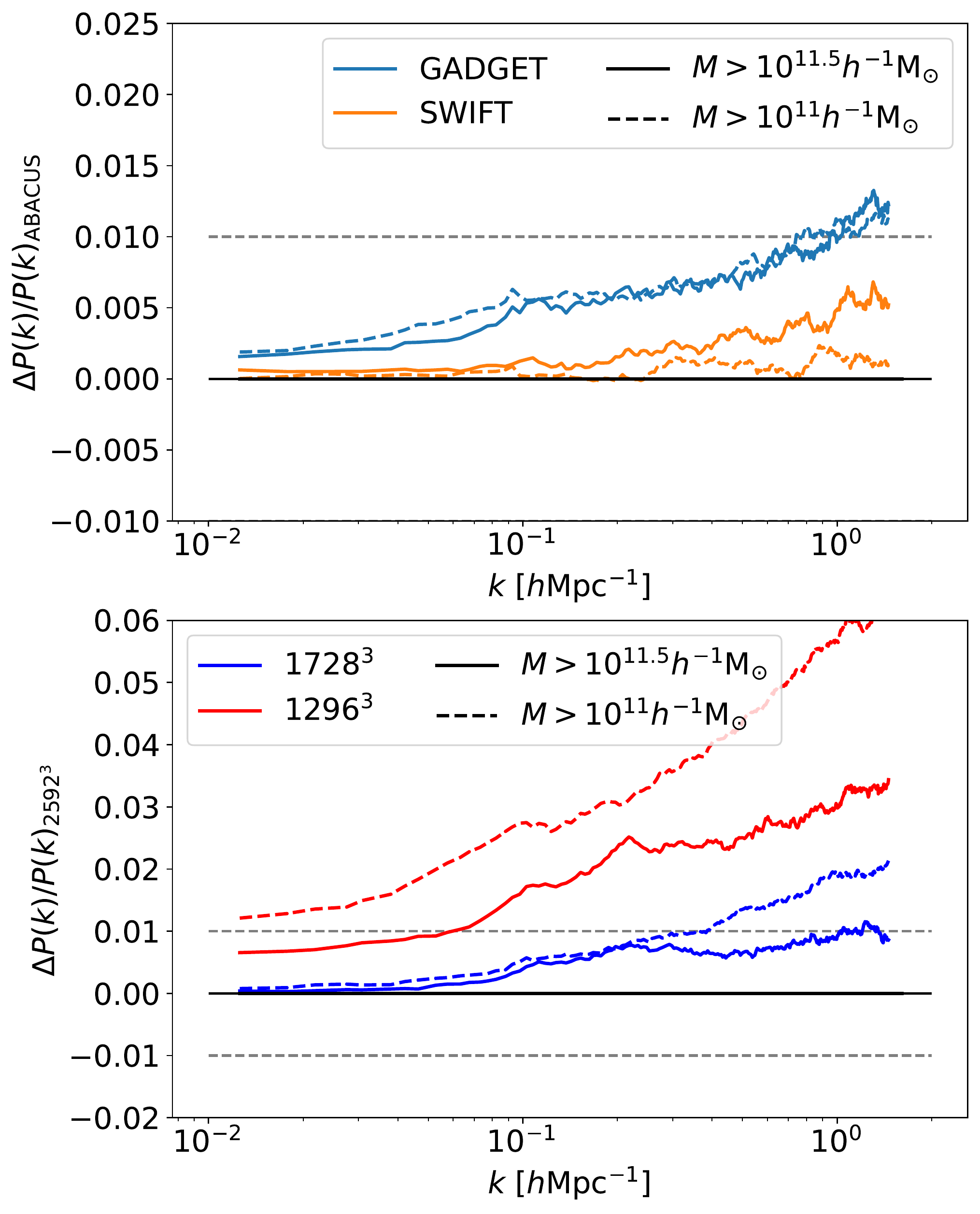}
    \caption{Upper panel: The ratio of the real-space power spectra of the halos from the high resolution simulations for different codes to \ABACUS at $z=1$ There is agreement between the codes to within the 1\% level at both mass cuts for $k < 1~h\, \rm{Mpc}^{-1}$, with \SWIFT and \ABACUS showing agreement to within 0.1\% for $k < 0.1~h\, \rm{Mpc}^{-1}$. Lower panel: The ratio of the real-space power spectra of the \ABACUS halos compared to the high resolution halo catalog. The medium resolution simulation agrees with the high resolution simulation more closely than the low resolution. The halo mass cut becomes more important for the resolution comparison than in the code comparison. The agreement becomes better with a higher halo mass cut in the resolution comparison.}
    \label{fig:pkr_both}
\end{figure}

\begin{figure}
	\includegraphics[width=\columnwidth]{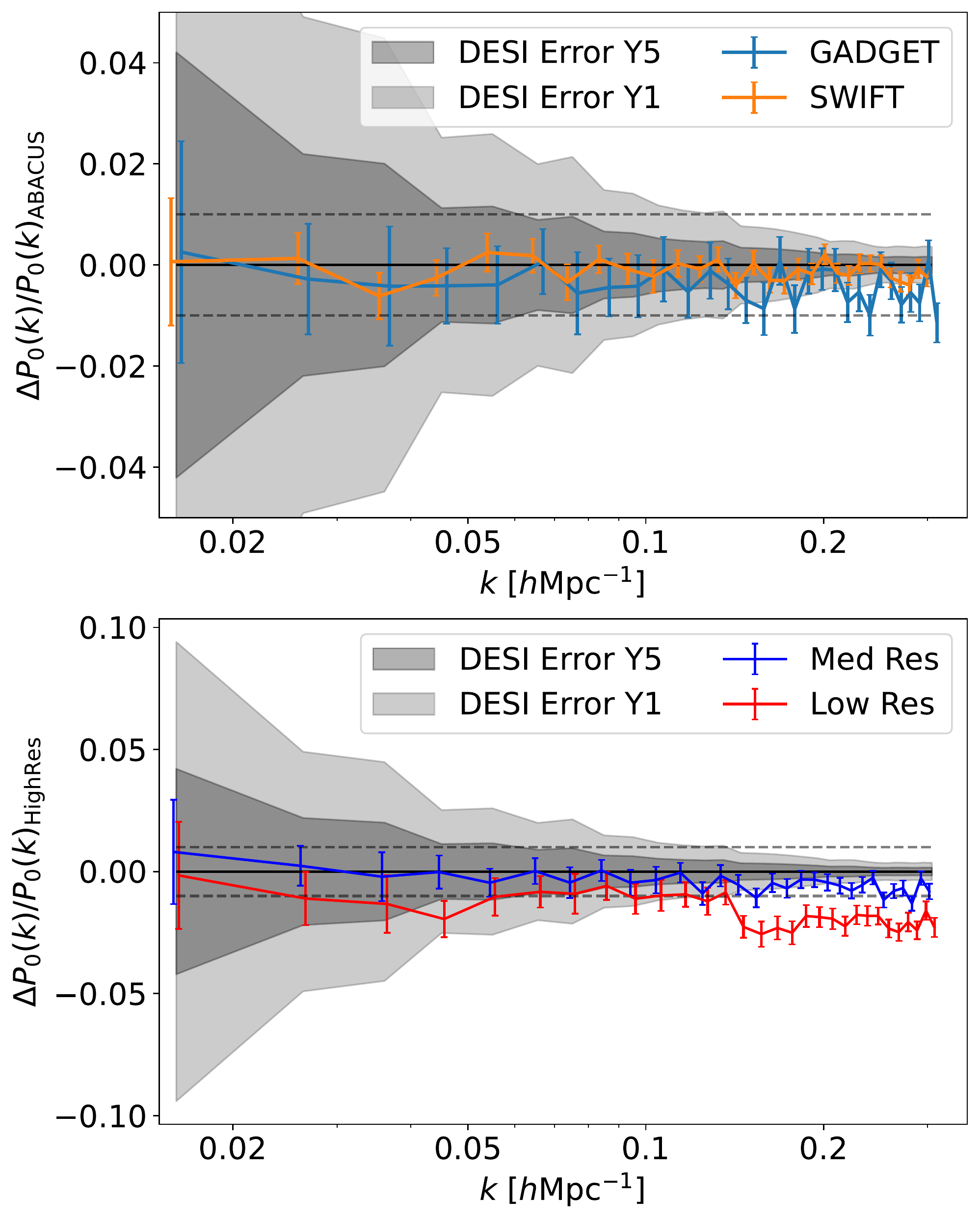}
    \caption{Upper panel: The ratio of the redshift space power spectrum monopoles of the halos from the high resolution simulations for different codes compared to \ABACUS at $z=1$. The shaded regions show an estimate for the DESI survey year 1 and year 5 statistical errors. The errorbars show an estimate for the noise while comparing two simulations which share the same phases. The codes show agreement to within the expected DESI statistical errors for $k < 0.1~h\, \rm{Mpc}^{-1}$. Lower panel: The ratio of the redshift space power spectrum monopoles from the high resolution simulations for \ABACUS at $z=1$. The medium resolution simulation is consistent with the high resolution simulation for $k < 0.2~h\, \rm{Mpc}^{-1}$. The low resolution simulation shows significant differences for $k > 0.15~h\, \rm{Mpc}^{-1}$.}
    \label{fig:pkl_0_both}
\end{figure}

\begin{figure}
	\includegraphics[width=\columnwidth]{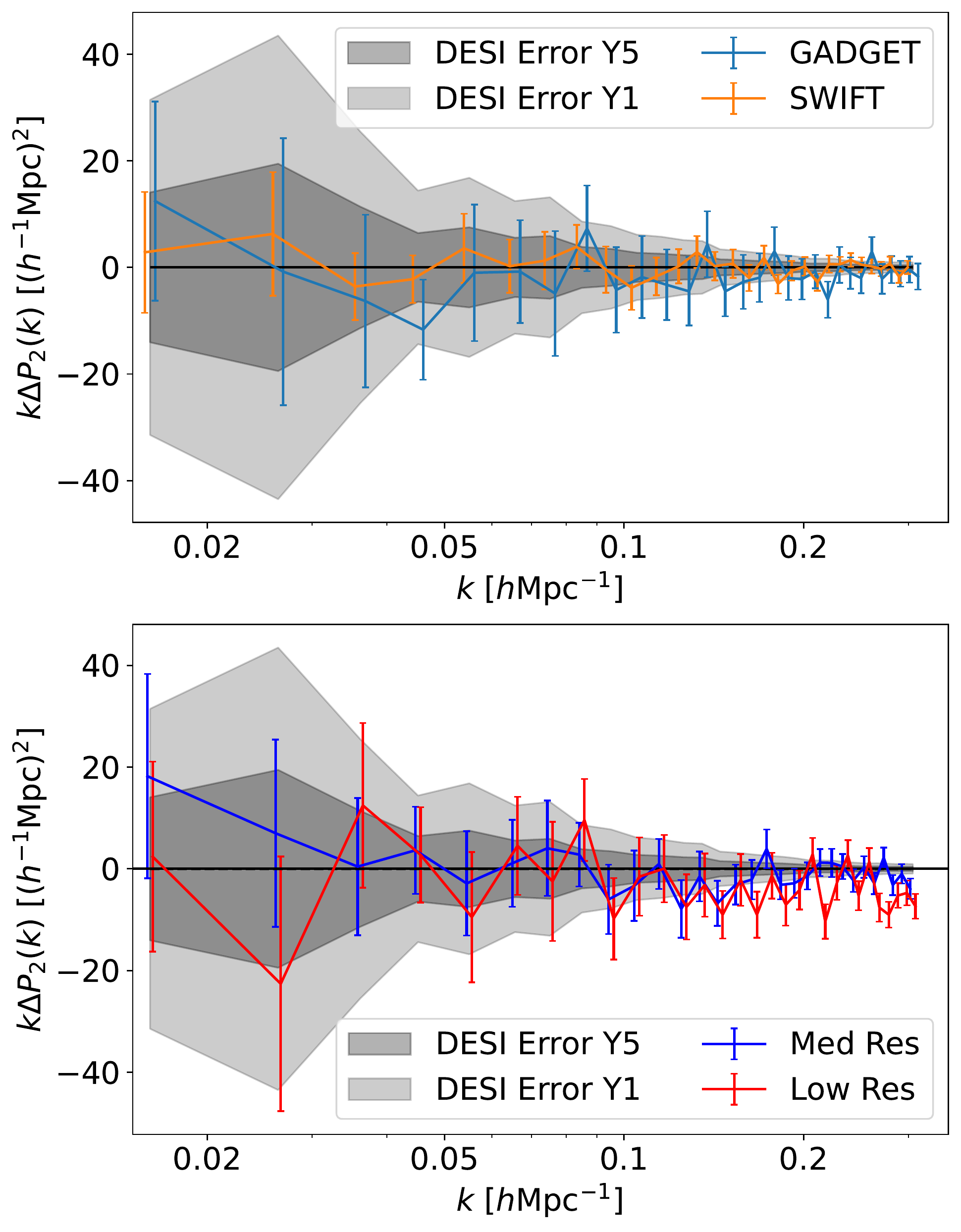}
    \caption{Upper panel: The difference of the redshift space power spectrum quadrupoles of the halos from the high resolution simulations for different codes compared to \ABACUS at $z=1$. The shaded regions show an estimate for the DESI survey year 1 and year 5 statistical errors. The errorbars show an estimate for the noise while comparing two simulations which share the same phases. Lower panel: The difference of the redshift space power spectrum quadrupoles from the high resolution simulations for \ABACUS at $z=1$. Due to the level of noise it is difficult to place bounds on the agreement of different codes and resolutions. For $k < 0.05~h\, \rm{Mpc}^{-1}$, where the simulation noise is comparable in magnitude to the DESI errors, the code and resolution comparisons appear to be consistent with one another.}
    \label{fig:pkl_2_both}
\end{figure}

Figs.~\ref{fig:pkr_both}\textendash\ref{fig:pkl_2_both} show the results from the halo power spectra comparisons between codes and resolutions, using two different halo mass cuts, in real-space and redshift space.

Jackknife errors are used to estimate the DESI survey statistical errors along with the noise due to finite simulation volume in a similar manner as in Section \ref{sec:Halo_Clustering}.
A detailed investigation of the jackknife method estimating the uncertainties in the power spectrum measurements will be presented in \cite{Zhang21}.

In Fig.~\ref{fig:pkr_both} the real-space power spectra are compared between codes. \SWIFT, \GADGET and \ABACUS agree at the 1\% level for $0.01\leq k  \leq 1$~$h\, \textrm{Mpc}^{-1}$. There is not a large difference between the power spectra for different halo mass cuts. The same effects are seen between the codes in the redshift space power spectrum monopole in Fig.~\ref{fig:pkl_0_both}, \SWIFT, \GADGET and \ABACUS are consistent within the DESI statistical errors up to $k = 0.2~h\, \rm{Mpc}^{-1}$.

Fig.~\ref{fig:pkr_both} shows that the medium and low resolution real-space halo power spectra are greater than the high resolution case. The difference increases at smaller scales and is larger for the lower halo mass cut and for the lower resolution simulation. In Fig.~\ref{fig:pkl_0_both} the same trends are seen in the redshift space monopole, with larger differences than the real-space power spectrum. The medium resolution simulation is consistent with the high resolution simulation for $k < 0.2~h\, \rm{Mpc}^{-1}$. The low resolution simulation shows significantly greater differences than the medium resolution simulation for $k > 0.15~h\, \rm{Mpc}^{-1}$.

The differences between the codes and resolutions in the redshift space quadrupole are shown in Figure ~\ref{fig:pkl_2_both}. Due to the level of noise it is difficult to draw conclusions about code and resolution agreement in the quadrupole. All the codes agree within the noise level, which is comparable to the DESI statistical error for $k < 0.05~h\, \rm{Mpc}^{-1}$. At smaller scales than this the simulation noise becomes larger than the expected DESI statistical error. 

\section{Conclusions}
\label{sec:Con}

We have performed two code comparison projects, comparing initial conditions generators and also $N$-body simulation codes. 

\subsection{IC Comparison}

The initial conditions code comparison included the IC generators associated with the \ABACUS, and \FASTPM $N$-body simulation codes, along with the IC generator \textsc{panphasia/ic\_gen}. ICs were generated at three different resolutions for each code. Figure \ref{fig:IC_compare} shows that the power spectra of the phase matched ICs from the same code agreed to within 1\% up to 20\% of the Nyquist frequency. At large scales the power spectra were consistent with the reference power spectrum, with any differences being within the expected variance due to finite volume. 

At small scales, \ABACUS, and \textsc{panphasia} agree with the input linear power spectrum within 1\% up to 50\% of the Nyquist frequency. Although we chose \textsc{panphasia} to setup the $N$-body code comparison project, the effect of the differences among the IC codes should be small.

The pairwise velocity anisotropy was used to check the consistency of the IC codes in the particle velocities as shown in Figure \ref{fig:IC_velocity}. The codes were mostly consistent with one another to within the sample variance estimated by using ten realisations of the \textsc{panphasia} ICs. \FASTPM showed larger pairwise anisotropy than the other codes at large scales.

\subsection{$N$-body simulation code comparison}

The $N$-body simulation codes, \ABACUS, \SWIFT, \GADGET, \& \FASTPM were run from identical \textsc{panphasia} ICs from $z=199$ to $z=0$. The matter power spectra were compared at $z=2$ and $z=1$. Comparisons between codes at fixed resolution were made, along with comparisons between resolutions for the same code.

When comparing between resolutions for each code in Figure \ref{fig:nbody_res_each_code}, the power spectra agree to within 1\% at $k = 2~h\,$Mpc$^{-1}$ and to within 4\% at $k = 10~h\,$Mpc$^{-1}$ with the exception of the quasi-$N$-body code \FASTPM which shows greater differences.
Comparing between the codes at fixed resolution the power spectra agree to within 0.1\% at large scales as seen in Figure \ref{fig:nbody_code_each_res}. Figure \ref{fig:perturbation} shows that the large scale power spectra are consistent with perturbation theory estimates made using 2LPT for all the codes. At small scales \SWIFT, \GADGET, and \ABACUS show good agreement, within 1\% at $k = 10~h\,$Mpc$^{-1}$ for all resolutions at both redshifts. As expected, \FASTPM shows the greatest difference in power spectrum relative to the other codes, disagreeing at the 1\% level for $k > 0.5~h\,$Mpc$^{-1}$. Similar conclusions are reached at the two explored redshifts.

The dark matter halos were found using \textsc{rockstar}. Halo properties were compared between matched halos for fixed code and variable resolution or fixed resolution and variable code and the results were shown in Figures \ref{fig:halo_prop_code} \& \ref{fig:halo_prop_resolution}. There was the greatest agreement between codes for high mass halos in most statistics. The difference between matched halo properties was within 1\% between all the codes for halo masses greater than $10^{13}$~$h^{-1}$ M$_{\odot}$.

Matching halo properties between different resolutions presented several systematic effects. Low mass halos in low resolution simulations had greater mass than the matched halos from the high resolution simulations, with the opposite effect observed at high masses. Matched halo properties had better agreement at high masses when comparing between simulations run with different resolutions.

The halo mass functions from \SWIFT, \GADGET, and \ABACUS agreed to within 1\% between mass limits of $10^{11.5} - 10^{14}$~$h^{-1}$ M$_{\odot}$ as seen in Figure \ref{fig:HMF_panel}.

Halo clustering measurements were made by using both the correlation function and power spectrum. Comparisons were made between codes and resolutions for two different halo mass limits $10^{11}$ and $10^{11.5}$~$h^{-1}$ M$_{\odot}$. The minimum halo mass that hosts the DESI ELG sample is expected to be between these two values. 

Comparing different codes in the halo clustering, all codes agree with each other within the expected DESI year 1 uncertainty at scales larger than $10\,h^{-1}$Mpc and within the DESI year 5 uncertainty at scales larger than $20\,h^{-1}$Mpc (Figs. \ref{fig:2PCF_both}-\ref{fig:xil_2_both}).
\ABACUS and \SWIFT have remarkable agreements down to $2\,h^{-1}$Mpc.
Comparing the different resolutions, the medium resolution run agrees with the high resolution one at scales larger than $20\,h^{-1}$Mpc. This indicates that one should be more careful choosing simulation resolution than choosing the $N$-body code.

In the halo power spectra, similar results to the halo correlation function were found as seen in Figs. \ref{fig:pkr_both}-\ref{fig:pkl_2_both}. \ABACUS, \SWIFT, and \GADGET have good agreement at all the scales which we are most interested in for DESI, i.e. $k < $0.3~$h\,$Mpc$^{-1}$. Medium resolution simulations agree with the high resolution simulations at a similar level to the code comparisons. The low resolution simulation shows differences at least twice as large as the medium resolution simulation at all length scales. The agreement becomes closer between different resolution simulations with a higher halo mass cut.

These results indicate the expected level of systematic errors in a variety of statistics associated with $N$-body simulations. 

We do not draw conclusions about whether certain simulations are appropriate for use in any specific DESI analyses, this requires further propagation of the simulations through the cosmology analysis pipeline. These simulations and the results provided should act as a baseline for expected systematic errors, regardless of the application of the simulations. Future work using these simulations will explore how these errors are propagated through HOD and cosmological analysis and therefore will assess the suitability of simulations run with different codes and resolutions for use in modern galaxy surveys \citep{HernandezAguayo21}.

\section*{Acknowledgements}
We thank Violeta Gonzalez-Perez, Graziano Rossi, Peter Behroozi, Joe DeRose (internal reviewer), and Sownak Bose (internal reviewer) for their constructive comments.
 This research is supported by the Director, Office of Science, Office of High Energy Physics of the U.S. Department of Energy under Contract No. DE–AC02–05CH11231, and by the National Energy Research Scientific Computing Center, a DOE Office of Science User Facility under the same contract; additional support for DESI is provided by the U.S. National Science Foundation, Division of Astronomical Sciences under Contract No. AST-0950945 to the NSF’s National Optical-Infrared Astronomy Research Laboratory; the Science and Technologies Facilities Council of the United Kingdom; the Gordon and Betty Moore Foundation; the Heising-Simons Foundation; the French Alternative Energies and Atomic Energy Commission (CEA); the National Council of Science and Technology of Mexico; the Ministry of Economy of Spain, and by the DESI Member Institutions.
The authors are honored to be permitted to conduct scientific research on Iolkam Du’ag (Kitt Peak), a mountain with particular significance to the Tohono O’odham Nation. 
CG is supported by a PhD Studentship from the Durham Centre for Doctoral Training in Data Intensive Science, funded by the UK Science and Technology Facilities Council (STFC, ST/P006744/1) and Durham University.
LHG is supported by the Center for Computational Astrophysics at the Flatiron Institute, which is supported by the Simons Foundation.  \ABACUS development has been supported by NSF AST-1313285 and DOE-SC0013718, as well as by Harvard University startup funds.
B.~L. would like to acknowledge the support of the National Research Foundation of Korea (NRF-2019R1I1A1A01063740) and the support of the Korea Institute for Advanced Study (KIAS) grant funded by the government of Korea. 
CH-A acknowledges support from the Excellence Cluster ORIGINS which is funded by the Deutsche Forschungsgemeinschaft (DFG, German Research Foundation) under Germany's Excellence Strategy - EXC-2094-390783311.
This work used the DiRAC@Durham facility managed by the Institute for Computational Cosmology on behalf of the STFC DiRAC HPC Facility (www.dirac.ac.uk). The equipment was funded by BEIS capital funding via STFC capital grants ST/K00042X/1, ST/P002293/1, ST/R002371/1 and ST/S002502/1, Durham University and STFC operations grant ST/R000832/1. DiRAC is part of the National e-Infrastructure.
The research in this paper made use of the \SWIFT open-source
simulation code (http://www.swiftsim.com, \citealt{2018ascl.soft05020S})
version 0.9.0.

\section*{Data Availability}

The authors make the initial conditions files, simulation snapshots, and halo catalogs available at a Globus link\footnote{\url{https://app.globus.org/file-manager?origin_id=53feea72-5215-11ec-a9cc-91e0e7641750&origin_path=\%2F}}. We encourage their use as a valuable resource for comparing the impact of $N$-body simulation code and resolution in other projects.

%%%%%%%%%%%%%%%%%%%%%%%%%%%%%%%%%%%%%%%%%%%%%%%%%%



\bibliographystyle{mnras}
\bibliography{example} % if your bibtex file is called example.bib

%\begin{thebibliography}{99}
%\end{thebibliography}

%%%%%%%%%%%%%%%%%%%%%%%%%%%%%%%%%%%%%%%%%%%%%%%%%%

\section*{Affiliations}
\noindent
{
\it% List of institutions
$^{1}$Institute for Computational Cosmology, Department of Physics, Durham University, Durham, DH1 3LE, UK\\
$^{2}$Kavli Institute for Particle Astrophysics and Cosmology, Stanford University, 452 Lomita Mall, Stanford, CA 94305, USA\\
$^{3}$Department of Physics and Astronomy, University of Utah, Salt Lake City, UT 84112, USA\\
$^{4}$Instituto de Astronomia, Universidad Nacional Autónoma de México, Apdo. Postal 20-364, México\\
$^{5}$Department of Physics, Manipur University, Canchipur, 795003, Manipur, India\\
$^{6}$Center for Computational Astrophysics, Flatiron Institute Simons Foundation, 162 Fifth Ave. New York, NY 10010, USA\\
$^{7}$Department of Physics and Astronomy, Sejong University, Seoul 05006, Korea\\
$^{8}$Berkeley Center for Cosmological Physics, Department of Physics, University of California Berkeley, Berkeley, CA 94720, USA\\
$^{9}$Max-Planck-Institut für Astrophysik, Karl-Schwarzschild-Str 1, D-85748 Garching, Germany\\
$^{10}$Excellence Cluster ORIGINS, Boltzmannstrasse 2, D-85748 Garching, Germany\\
$^{11}$Institute for Astronomy, University of Edinburgh, Royal Observatory, Blackford Hill, Edinburgh EH9 3HJ, UK\\
$^{12}$Department of Physics, Kansas State University, Manhattan, KS 66506, USA\\
$^{13}$Department of Astronomy, School of Physics and Astronomy, Shanghai Jiao Tong University, Shanghai 200240, People's Republic of China\\
$^{14}$Center for Astrophysics | Harvard \& Smithsonian, 60 Garden St., Cambridge, MA 02138, USA\\
$^{15}$Centre for Extragalactic Astronomy, Department of Physics, Durham University, Durham, DH1 3LE, UK\\
$^{16}$University College London, Department of Physics \& Astronomy, Gower Street, London, WC1E 6BT \\
$^{17}$Lawrence Berkeley National Laboratory, 1 Cyclotron Road, Berkeley, California 94720, USA\\
$^{18}$NSF's National Optical-Infrared Astronomy Research Laboratory, 950 N. Cherry Ave., Tucson, AZ 85719, USA\\
$^{19}$Space Sciences Laboratory (SSL), UC Berkeley\\
$^{20}$Physics Department, University of Michigan Ann Arbor, MI 48109, USA}

%%%%%%%%%%%%%%%%% APPENDICES %%%%%%%%%%%%%%%%%%%%%

\appendix

\section{Comparison of Subhalo Mass Functions}

\begin{figure}
	\includegraphics[width=\columnwidth]{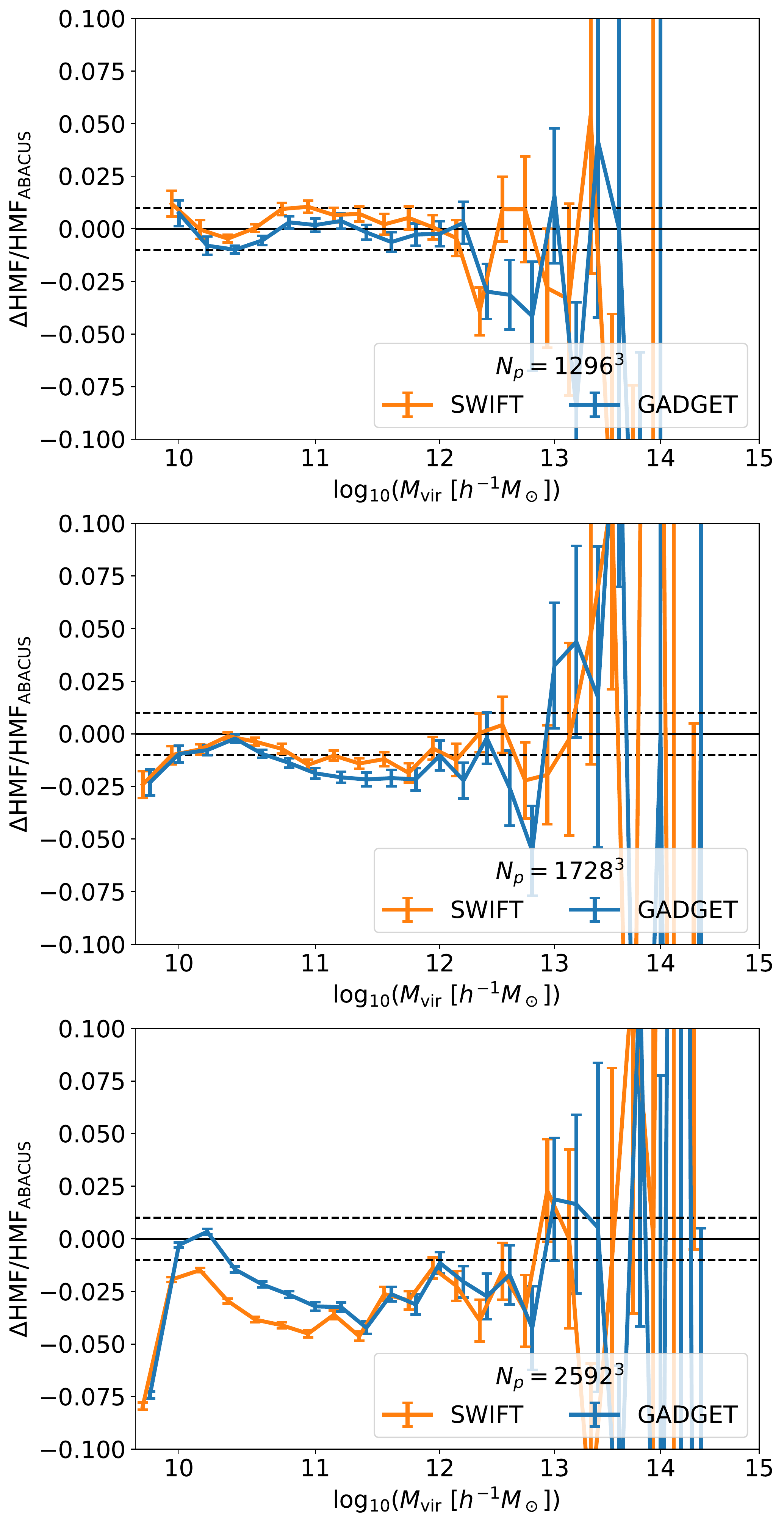}
    \caption{The ratio of the subhalo mass functions to that of the \ABACUS simulation at fixed resolution at $z=1$. The top, middle and bottom panels show the low, medium, and high resolution simulations respectively. The errorbars show an estimate for the noise while comparing two simulations which share the same phases.}
    \label{fig:hmf_subhalo}
\end{figure}

The comparison of subhalo mass functions between the simulations at fixed resolution at $z=1$ is presented in Figure \ref{fig:hmf_subhalo}. The differences between the subhalo mass functions are smaller at lower resolution, lying within 1\%, 2\%, and 5\% in the low, medium, and high resolutions respectively between $10^{11} \textrm{and} 10^{12}~h^{-1}$ M$_{\odot}$. \SWIFT and \GADGET show good agreement within 1\% for all resolutions between the mass ranges of $10^{11} \textrm{and} 10^{12}~h^{-1}$ M$_{\odot}$. Above $10^{13}~h^{-1}$ M$_{\odot}$ the subhalo mass functions become noisy. The subhalo mass function is likely to be more sensitive to differences in the softening schemes of the different simulations than the parent halo mass function.

%%%%%%% %%%%%%%%%%%%%%%%%%%%%%%%%%%%%%%%%%%%%%%%%%%

% Don't change these lines
\bsp	% typesetting comment
\label{lastpage}
\end{document}